\documentclass[epj]{svjour}
\usepackage{graphics}
\usepackage{hepunits2007}
\usepackage{lhc_defs}

\usepackage{calc}
\usepackage{bm}
\usepackage{makeidx}
\usepackage{subfigure}
\usepackage{xspace}
\usepackage[noprefix]{nomencl}
\usepackage{marvosym}
\usepackage{wasysym}
\usepackage{delarray}
\usepackage{footmisc}
\usepackage{graphicx}
\usepackage[sectionbib,numbers,sort,compress]{natbib}
\usepackage{color}
\usepackage[hyperfootnotes=false]{hyperref}
\graphicspath{{.}}

\begin{document}

\title{The Status of Supersymmetry after the LHC Run 1}
\author{Philip Bechtle$^{\dagger}$, Tilman Plehn$^{\ddagger}$ and Christian Sander$^{\amalg}$}
\institute{$^{\dagger}$Physikalisches Institut, Universit\"at Bonn, Nussallee~12, 53115~Bonn, Germany\\$^{\ddagger}$Institut f\"ur Theoretische Physik, Universit\"at Heidelberg, Philosophenweg~16, 69120~Hamburg, Germany\\$^{\amalg}$Institut f\"ur Experimentalphysik, Universit\"at Hamburg, Luruper Chaussee~149, 22761~Hamburg, Germany}

\abstract{Supersymmetry (SUSY) is a complete and renormalisable candidate
      for an extension of the Standard Model.
      At an energy scale not too far above the electroweak scale it would solve
      the hierarchy problem of the SM Higgs boson, dynamically explain
      electroweak symmetry breaking, and provide a dark-matter
      candidate. Since it doubles the Standard Model degrees of
      freedom, SUSY predicts a large number of additional particles,
      whose properties and effects on precision measurements can be
      explicitly predicted in a given SUSY model.
      In this review the motivation for SUSY is outlined, the various searches strategies for SUSY particles at the LHC are described, and
      the status of SUSY in global analyses after the LHC Run 1 is summarized.}

\authorrunning{Bechtle, Plehn, Sander}
\titlerunning{The Status of Supersymmetry after the LHC Run 1}

\maketitle
\tableofcontents
\blfootnote{Originally published in ``The Large Hadron Collider --- Harvest of Run~1'', edited by T.~Sch\"orner-Sadenius, Springer, 2015, pp. 421--462~\cite{thisbook}.}

%%%%%%%%%%%%%%%%%%%%%%%%%%%%%%%%%%%%%%%%%%%%%%%%%%%%%%%%%%%%%%%%%%%
\section{A Short Motivation}\label{sec:susy:intro}

In many ways, the Standard Model (SM)
\index{Standard Model} 
including a Higgs
boson~\cite{Higgs:1964ia,Higgs:1964pj,Englert:1964et,Chatrchyan:2012ufa,Aad:2012tfa} is a complete theory, both from an
experimental and from a theoretical point of view. It explains all
observations at the LHC and at preceding high-energy experiments and
shows no conceptual inconsistencies. Even better, it is a
renormalisable field theory with all gauge couplings remaining
perturbative to the Planck scale $\MP \sim \unit{\power{10}{19}}{\GeV}$.
Aside from gravity---which is missing in the Standard Model---we could 
be happy with this rather complete picture after the Higgs discovery~\cite{Plehn:2015dqa}. However, if the perturbatively renormalisable Standard Model 
is supposed to describe
all measurements in particle physics over such a wide range of energies,
it looks much less complete~\cite{Martin:1997ns,Morrissey:2009tf}.

If we widen our experimental horizon beyond LHC energies, several
shortcomings
%\index{Standard Model!shortcomings} 
arise: 
Most notably, it does not
explain dark matter~\cite{Jungman:1995df,Bertone:2004pz}; it does not
accommodate the matter-antimatter asymmetry of the
universe~\cite{Bernreuther:2002uj}; and it gives no reason  
why the three gauge couplings of the Standard Model gauge group almost meet at
a large unification scale, but not quite~\cite{Amaldi:1991cn}.\index{dark matter}\index{matter-antimatter asymmetry}\index{unification}

Gauge coupling unification
can be viewed as an experimental motivation for physics
beyond the Standard Model, but it is closely tied to a list of open
theoretical questions: Where does the flavour structure in the quark
and lepton sectors come from? Is there a high-scale simplification of
the Standard Model gauge structure? Why is the Higgs mass so much
smaller than the Planck scale, which we consider 
the ultimate physical cut-off scale of the
theory~\cite{Gildener:1976ai,Weinberg:1975gm,Lee:1977yc}? 

As the only fundamental scalar particles in the Standard Model,
the newly discovered Higgs boson causes peculiar
problems in its perturbative field theory description.  In the
presence of a physical cut-off scale, its mass turns out to be extremely
sensitive to quantum corrections.  This means that we would expect
such a scalar to acquire a mass close to the cut-off scale. On the
other hand, electroweak symmetry breaking and the unitarity of \WWnosign
scattering require the Higgs mass to stay sufficiently below the \TeV\ 
scale. This tension is called the ``gauge hierarchy problem''
%\index{gauge hierarchy problem|see{hierarchy problem}} 
or ``hierarchy problem''.
\index{hierarchy problem}

Supersymmetry is a leading candidate for a theory that can solve many
of these problems~\cite{Golfand:1971iw,Volkov:1972jx,Wess:1974tw,Salam:1974yz}. It
is based on an extension of the symmetry structure of the Standard
Model by a symmetry linking fermions and bosons, i.e.\ matter fields
and force carriers~\cite{Haag:1974qh}. In that sense, supersymmetry completes the
concept of symmetries as the guiding principle of elementary particle
physics. While models with more than one supersymmetry generator are
useful tools to study the structure of gauge theories, 
we will focus on the phenomenologically most interesting case of $\mathcal{N}=1$
supersymmetry~\cite{Wess:1992cp,Terning:2006bq}.

This supersymmetric version of the Standard
Model~\cite{Martin:1997ns,Drees:2004jm,Baer:2006rs} protects the Higgs-boson
mass from quadratic divergences. It improves the unification of the
three gauge couplings and moves the gauge-unification scale to values
around $M_\text{GUT} \simeq \unit{\power{10}{16}}{\GeV}$, large enough to avoid
predicting unobserved proton decays~\cite{Dawson:1979zq,Einhorn:1981sx,Dimopoulos:1981zb,Dine:1981gu}.
It can provide a weakly interacting dark-matter candidate, and it might
help explaining the matter-antimatter asymmetry.\index{matter-antimatter asymmetry} And what makes it
most relevant to LHC physics: It predicts new fundamental particles at
the \TeV\ scale.

%%%%%%%%%%%%%%%%%%%%%%%%%%%%%%%%%%%%%%%%%%%%%%%%%%%%%%%%%%%%%%%%%%%
\section{Theoretical Introduction}
\label{sec:susy:theory}

The minimal supersymmetric Standard Model (MSSM)
%\index{MSSM|see{supersymmetry!minimal supersymmetric Standard Model}}
%\index{minimal supersymmetric Standard Model|see{supersymmetry}}
%\index{supersymmetry!minimal supersymmetric Standard Model}
\index{SUSY!MSSM}
is the simplest supersymmetric extension of the Standard Model.
Through the additional symmetry requirement it first of all doubles the number of fundamental particle fields.  Since its
underlying symmetry respects the number of degrees of freedom, it
predicts two complex scalar fields for each Standard Model fermion,
which can be arranged to match the two helicities of the Dirac
field. For each Standard Model fermion $\ensuremath{f}$ we denote
these two scalar ``sfermions''
%\index{sfermion}
as $\sfermion_{L,R}$.
The gauge bosons get mapped onto fermionic fields, where the massless
photon and gluon fields only provide enough degrees of freedom for
Majorana partners. This means that the discovery e.g.\ of a strongly
interacting Majorana gluino $\gluino$ 
%\index{gluino} 
would be a clear sign for an
underlying supersymmetry.

%%%%%%%%%%%%%%%%%%%%%%%%%%%%%%%%%%%%%%%%%%%%%%%%%%%%%%%%%%%%%%%%%%%
\subsection{Minimal Supersymmetric Standard Model}\label{sec:susy:theory:mssm}

Two additional aspects of the MSSM that are of crucial relevance to
LHC searches are also directly linked to the properties of
supersymmetric gauge theories.

First, the so-called superpotential, written in terms of superfields
with their fermionic and bosonic constituent fields, predicts
interactions where (scalar) supersymmetric partner states of quarks (squarks) 
%\index{squarks}
and leptons (sleptons) 
%\index{sleptons} 
are coupled to two Standard Model fermions.  These
interactions break lepton number and baryon number and mediate rapid
proton decay. The easiest way to remove them is by postulating a new
symmetry or multiplicative quantum number, called $R$-parity. 
\index{R-parity@$R$-parity}  
All
Standard Model states are assigned $R=+1$, which means they transform
even under an $R$-transformation. In contrast, supersymmetric partner
states with $R=-1$ are odd.  
$R$-parity thus requires that
any interaction vertex include an even number of supersymmetric
partner states~\cite{Farrar:1978xj,Dreiner:1997uz}. If, for example, we
consider supersymmetric QCD with $R$-even quark and gluon fields and
$R$-odd (scalar) quark partners \squark and (fermionic) gluon partners
\gluino, the allowed supersymmetric 3-point gauge interactions are
%
%\begin{linenomath}
\begin{equation*}
\qq\qq\glue, \; 
\glue\glue\glue
\quad 
\stackrel{\text{SUSY}}{\longrightarrow}
\quad
\squark \squark \glue, \; 
\qq \squark \gluino, \;
\gluino \gluino \glue \; .
\end{equation*}
%\end{linenomath}
%
Three-squark interactions do not exist because they would lead to
proton decay and need to be forbidden. In addition, $R$-parity has a
crucial effect on the ``lightest supersymmetric partner''
(LSP):
%\index{LSP|see{lightest supersymmetric partner}}
%\index{lightest supersymmetric partner|see{supersymmetry}}
\index{SUSY!LSP}
As an $R$-odd state, it cannot decay to two
Standard Model particles. Since, however, there is also no lighter supersymmetric
partner to which it could decay through an allowed interaction, it has to be
stable and can therefore serve as a dark-matter candidate.

Second, because of the underlying supersymmetry, the superpotential has
to be holomorphic, i.e.\ it cannot include a superfield and its
complex conjugate at the same time. This property directly translates
into the Higgs scalar as a constituent of the superfield.  The Higgs sector in the Standard
Model gives mass to up-type and down-type fermions through the Higgs
field and its conjugate.  In the MSSM this is not possible, which
means that the Standard Model ($R=+1$) sector of the MSSM has to be
modified~\cite{Gunion:1989we,Djouadi:2005gj}: Instead of one Higgs
doublet we need at least two doublets to provide masses to all fermion
fields. In the minimal setup of the MSSM, these doublets are arranged
such that at tree level one 
of them
couples to up-type fermions and the other
to down-type fermions. In combination, their two vacuum expectation
values have to provide the weak gauge boson masses, $v_u^2 + v_d^2 =
v^2 = \left( \unit{246}{\GeV} \right)\squared$.  Two complex Higgs
doublets are built of eight degrees of freedom.  Just as in the
Standard Model, one pseudo-scalar degree of freedom and one set of
charged degrees of freedom become the Goldstone modes of the massive
\Zb and \Wpm bosons.\index{Goldstone bosons}  The remaining five degrees of freedom, including
a pseudoscalar and a charged scalar, turn into physical Higgs fields:
two scalar Higgs fields $H_1^0$ and $H_2^0$, one pseudo-scalar Higgs
$A^0$, and a pair of 
charged Higgs fields $H^\pm$.  The two scalar states mix to
mass eigenstates, $h^0$ and $H^0$. As will be discussed below, the
mass of the lightest supersymmetric Higgs boson is bounded from above.

In particular the structure of scalar interactions in the MSSM is not
obvious from the Standard Model Lagrangian. The reason is that
supersymmetry respects the number of degrees of freedom not only of
the on-shell particles, but also of the off-shell fields. Already in
the simple Wess--Zumino
model~\cite{Wess:1974tw}
%\index{SUSY!Wess--Zumino model} 
with
its chiral (matter) superfield, we observe that the conservation of
off-shell degrees of freedom as well as the consistency of
supersymmetry transformations staying within each superfield require
an auxiliary field $F$. It does not feature a kinetic term and can
therefore be removed through its equation of motion. After integrating
out the $F$ field, this structure ensures that the Standard Model
matter fermions and their scalar partners have the same masses.
If we include gauge interactions via a vector superfield this links the
Standard Model gauge bosons to their fermionic gaugino partners and
leads to another auxiliary field, $D$. This field has gauge couplings
to scalars of the kind $D \phi^\dagger \phi$, which after exploiting
the equation of motion introduces gauge couplings of the type
$\phi^4$. In the Higgs sector they replace the free quartic coupling
$\lambda \phi^4$ and hence determine the masses of the supersymmetric
Higgs bosons.  After diagonalising the Higgs eigenstates, we can
express the mass of the lightest Higgs boson 
in terms of the ratio of the two
vacuum expectation values $\tan \beta = v_u/v_d$,
namely~\cite{Chankowski:1992er,Haber:1993an,Hempfling:1993qq,Carena:2000dp}
%
%\begin{linenomath}
\begin{equation}
\Mh = \MZ \cos (2 \beta) 
    + \text{radiative corrections} < \unit{135}{\GeV} \, ,
\label{eq:susy:higgsmass}
\end{equation}
%\end{linenomath}
%
where the limit assumes squarks in the \TeV\ range.  This is one of
the key predictions of the MSSM and is based on the structure of its
scalar potential. In the light of the experimental constraints
discussed in \sect{\ref{sec:susy:status:Fits}}, the additional MSSM
Higgs states should be significantly heavier.  

%-------------------------------------------------------
\begin{table}[t]
\begin{center}
\begin{tabular}{|l|c|r|r|r|}
\hline
State&Symbol&$\phantom{i3}R\phantom{i3}$&$\text{SU}(3)_{\text{c}}$&$\phantom{3}Q_\text{EM}\phantom{3}$\\
\hline
\CP-even Higgs  & $h^0,H^0$ & $+1$ & 1 & 0\\
\CP-odd Higgs & $A^0$ & $+1$ & 0 & 0\\
charged Higgs & $H^\pm$ & $+1$ & 1 & $\pm1$\\
\hline
gluino & $\gluino$ & $-1$ & 8 & 0 \\
neutralinos & $\tilde{\chi}_1^0, \tilde{\chi}_2^0, \tilde{\chi}_3^0, \tilde{\chi}_4^0$ & $-1$ & 1 & 0\\
charginos & $\tilde{\chi}^\pm_1, \tilde{\chi}^\pm_2$ & $-1$ & 1 & $\pm 1$\\
\hline
up squarks & $\tilde{\uq}_{L,R}, \tilde{\cq}_{L,R}, \tilde{\tq}_{1,2}$ & $-1$ & 3 & 2/3\\
down squarks & $\tilde{\dq}_{L,R}, \tilde{\sq}_{L,R}, \tilde{\bq}_{1,2}$ & $-1$ & 3 & $-1/3$\\
sleptons & $\tilde{e}_{L,R}, \tilde{\mu}_{L,R}, \tilde{\tau}_{1,2}$ & $-1$ & 1 & $\pm 1$\\
sneutrinos & $\tilde{\nu}_{e}, \tilde{\nu}_{\mu}, \tilde{\nu}_{\tau}$ & $-1$ & 1 & 0\\
\hline
gravitino & $\tilde{G}$ & $-1$ & 1 & 0 \\
\hline
\end{tabular}
\caption{MSSM mass eigenstates after electroweak symmetry breaking. We
  assume that left-right scalar mixing is only relevant for the
  massive third generation.
\label{tab:susy:particles}}
\end{center}
\end{table}
%-------------------------------------------------------

If we add supersymmetric partners for the entire Standard Model spectrum, 
keeping in mind that the two Higgs doublets are $R$-even, we find the MSSM particle
content
%\index{supersymmetry!particle content} 
shown in
\tab{\ref{tab:susy:particles}}. As mentioned before, every Standard
Model fermion acquires a scalar partner, called squark, slepton or
sneutrino.  The first-generation and second-generation sfermions with
their negligible fermion-Higgs couplings can, to a good approximation,
be represented by their interaction states, while the heavy-flavour
sfermions have to be rotated into their mass eigenstates. This follows
from the structure of their masses and mixings, which is determined by
supersymmetry breaking.

Every Standard Model boson matches onto a fermionic partner; one
example is the gluino 
%\index{gluino} 
as the partner of the gluons. 
Because supersymmetry respects the number of degrees of freedom, an on-shell
gluino without counting the colour structure can only be made out of two degrees of freedom. This does not
allow us to define a Dirac fermion and leaves us with a Majorana
fermion, which forms its own anti-particle.  In the interaction basis
the partner states of the electroweak gauge bosons and the extended
Higgs sector occur as one neutral bino, 
%\index{bino} 
charged and neutral winos, 
%\index{wino} 
and
charged and neutral Higgsinos. 
%\index{higgsino} 
These states can mix into mass
eigenstates, called neutralinos 
%\index{neutralino} 
and charginos 
%\index{chargino} 
and ordered by mass.
The neutralinos, just like the gluino, are Majorana fermions, as
suggested by the available two on-shell degrees of freedom. The
charginos with their electric charge cannot be their own
anti-particles, so they instead form Dirac fermions with interactions
that violate particle number.  Finally, the gravitino 
%\index{gravitino} 
as the spin-3/2
partner of the graviton can play a role at the LHC in specific
supersymmetric models.

%%%%%%%%%%%%%%%%%%%%%%%%%%%%%%%%%%%%%%%%%%%%%%%%%%%%%%%%%%%%%%%%%%%
\subsection{Supersymmetry Breaking}
\label{sec:susy:theory:breaking}

Obviously, supersymmetry is not an exact symmetry of nature, because
we have not observed any scalar partners of the Standard Model
fermions with the same masses. Without affecting the stabilisation of
the quadratically divergent Higgs mass and the improved gauge coupling
unification, we can introduce masses for all supersymmetric partners
through so-called soft breaking~\cite{Girardello:1981wz}.\index{unification}
\index{SUSY!soft SUSY breaking}
In this way, the supersymmetric partners receive two contributions to their
masses from electroweak symmetry breaking as well as from soft
supersymmetry breaking.  
Such soft-breaking masses remove the quadratic
divergence of the Higgs mass, but leave a logarithmic dependence. 
The largest contribution to the light Higgs mass is proportional to the
logarithm of the top-stop mass ratio, $\log (\Mstop/\Mt)$.  This
logarithmic mass dependence 
leads to the dominant radiative
corrections in \eqn{\eqref{eq:susy:higgsmass}} and is referred to as
the ``little hierarchy problem''
\index{hierarchy problem} 
of
supersymmetry. It is a key ingredient to so-called
naturalness
\index{naturalness} 
arguments. In general, naturalness
describes to what degree we have to tune ratios of model parameters
in order to accommodate experimental constraints 
on a model. If the light Higgs mass given in \eqn{\eqref{eq:susy:higgsmass}}
is considered a prediction over the MSSM parameter space, 
a large stop mass 
moves us away from the supersymmetric limit $\Mt = \Mstop$ and is
hence fine-tuned or un-natural.  While a discussion of these effects
will be part of a theoretical analysis of a given supersymmetric model
after its discovery, its benefit in guiding LHC analyses is not at all
obvious. 

Including the soft breaking terms, the relevant Lagrangian of the
MSSM
\index{SUSY!MSSM}
reads~\cite{Martin:1997ns}
%
%\begin{linenomath}
\begin{alignat}{5}
-\mathcal{L}_\text{soft} 
&= m_{H_d}^2 |H_d|^2 +  m_{H_u}^2\,|H_u|^2
  -(B\mu\,H_u\cdot H_d + \text{h.c.})\notag \\
&  + m_{Q}^2\,|\tilde{Q}|^2
   + m_{U}^2\,|\tilde{U}^{c}|^2
   + m_{D}^2\,|\tilde{D}^{c}|^2 
   + m_{L}^2\,|\tilde{L}|^2
   + m_{E}^2\,|\tilde{E}^{c}|^2 \notag \\
&  + \left(y_uA_u\,\tilde{Q}\cdot H_u\,\tilde{U}^c 
   -  y_dA_d\,\tilde{Q}\cdot H_d \tilde{U}^c
   - y_eA_e\,\tilde{L}\cdot H_d\tilde{E^c} + \text{h.c.} \right) \notag \\
&  +\left( \frac{M_3}{2}\,\tilde{g}^a\tilde{g}^a 
   + \frac{M_2}{2} \,\tilde{\Wb}^d\tilde{\Wb}^d
   + \frac{M_1}{2} \,\tilde{B}^0\tilde{B}^0 + \text{h.c.} \right) \; ,
\label{eq:susy:breaking:soft}
\end{alignat}
%\end{linenomath}
% 
where we omit flavour indices.  
The first line contributes to the
Higgs potential, as do the supersymmetric $F$ and $D$ terms described
above.  The second and third lines generate masses and mixings for the
sfermions. The couplings of the kind $y_j A_j$ describe left-right
mixing in the Higgs interaction and turn into mixing terms in the
squark-mass matrix after electroweak symmetry breaking.  Assuming that
they are proportional to the fermion Yukawa interaction ensures that
the one-loop contributions to the fermion masses are protected by the chiral symmetry, i.e.\
they scale with the fermion masses themselves.\index{chiral symmetry} 
In \tab{\ref{tab:susy:particles}} we neglect them for the first two
generations. The final line in the Lagrangian \eqref{eq:susy:breaking:soft} 
contains three gaugino Majorana mass terms.

If we expand the Lagrangian \eqref{eq:susy:breaking:soft} in
its most general form, the number of free parameters exceeds 100,
which makes the MSSM scenario somewhat unpredictive. This parameter
set includes complex gaugino masses and trilinear mixing parameters,
which are strongly constrained by electric dipole moments, or a
general flavour sector with 6$\times$6 matrices for the squark
masses, clearly not reflecting the observed flavour symmetries. For
direct LHC searches with their given energy range and their limited
precision, we can reduce this set to the LHC-relevant parameters,
usually referred to as the phenomenological MSSM
(pMSSM).
%\index{pMSSM|see{phenomenological MSSM}}
%\index{phenomenological MSSM|see{supersymmetry}}
\index{SUSY!pMSSM}
Its specific assumptions are:
no $R$-parity violation, no new sources of \CP violation,
mass-degenerate first and second generation scalars, and no
flavour-changing neutral currents. These assumed symmetry properties
automatically avoid a large number of indirect constraints on the
supersymmetric Lagrangian and allow for a dark-matter candidate.  They
leave us with 19 supersymmetric model parameters at the \TeV\ scale:
\begin{itemize}
  \item[] Higgs potential parameters: $\tan \beta$, $\mu$, and $\MA$\,;
  \item[] gaugino masses: $M_1$, $M_2$, and $M_3$\, ;
  \item[] squark mass parameters: $m_{\tilde{\tq}_L}$,$m_{\tilde{\tq}_R}$,$m_{\tilde{\bq}_R}$,
                                  $m_{\tilde{\uq}_L}$,$m_{\tilde{\uq}_R}$, and $m_{\tilde{\dq}_R}$\, ;
  \item[] slepton and sneutrino mass parameters: $m_{\tilde{\tau}_L}$,$m_{\tilde{\tau}_R}$,
                                                 $m_{\tilde{e}_L}$, and $m_{\tilde{e}_R}$\, ;
  \item[] trilinear couplings: $A_t$, $A_b$, and $A_\tau$\, .
\end{itemize}
In the pMSSM, the lightest neutralino is the dark-matter
candidate~\cite{Goldberg:1983nd,Ellis:1983ew}, because the only other
weakly interacting candidate, sneutrino dark matter, is ruled out
experimentally. The gravitino only couples
gravitationally to the rest of the spectrum.  As long as it is not the
LSP, it will therefore not play a role in LHC phenomenology. However,
if it should be the LSP and lead to decays of long-lived particles, it
can easily be added to the pMSSM.

At the \TeV\ scale, all pMSSM parameter can be chosen
independently. However, if we attempt to link these model parameters
to the structure of supersymmetry breaking, we will most likely evolve
them to higher energy scales, above which supersymmetry will remain
unbroken. Under renormalisation group evolution the squark and gluino
masses behave differently: The gluino mass only receives
multiplicative corrections, while the squark mass gets corrected by
terms proportional to the gluino mass. In this way a heavy gluino will
induce large squark masses, which means that we cannot link models
with $m_{\tilde{g}} \gg m_{\tilde{\qq}}$ to high-scale
theories~\cite{Jaeckel:2011wp}. To a lesser extent the same is true
for sleptons and electroweak gauginos. From a phenomenological
perspective this does not cause any problems, because indirect
constraints mostly require heavy scalars. The extreme case of
decoupled sfermions is referred to as ``split
supersymmetry''
~\cite{ArkaniHamed:2004fb,Giudice:2004tc},
and will be discussed in \sect{\ref{sec:susy:Exoticsearches:LongLived}}.
%\index{split SUSY|see{supersymmetry}}
\index{SUSY!split SUSY}

Historically, many searches for supersymmetry were not only
interpreted, but also set up with a specific SUSY-breaking mechanism
in mind.  Actual supersymmetry breaking can be described by the
auxiliary $F$ and $D$ fields introduced in
\sect{\ref{sec:susy:theory:mssm}}.  
However,
for LHC phenomenology supersymmetry breaking is simply an experimental fact;
what is more relevant is the way supersymmetry breaking is mediated to
the \TeV-scale Lagrangian.  
%hereherehere
The most popular mechanism is gravity mediation,
%\index{gravity mediation|see{supersymmetry!gravity mediation}}
%\index{supersymmetry!gravity mediation}
either
referred to as a supergravity-inspired model (mSUGRA)
%\index{mSUGRA|see{supersymmetry!mSUGRA}}
\index{SUSY!mSUGRA}
or as a strongly
constrained minimal supersymmetric model (CMSSM).
%\index{CMSSM|see{constrained MSSM}}
%\index{constrained MSSM|see{supersymmetry}}
\index{SUSY!constrained MSSM}  
In these models the sector communicating SUSY breaking does not couple to the gauge
structure of the Standard
Model~\cite{Ibanez:1981yh,Chamseddine:1982jx,Barbieri:1982eh}. This
requires all gaugino masses, all scalar mass parameters 
(of Higgs bosons, squarks and sleptons) 
and all trilinear couplings to unify at the
same scale. The CMSSM model parameters are $m_{1/2}$, $m_0$, and
$A_0$, all defined at $M_\text{GUT}$. In addition, we specify $\tan
\beta$ and the sign of $\mu$ in the Higgs sector. Strictly speaking,
in the corresponding mSUGRA model 
the free model parameter is $B = A_0 -
m_0$.  
After renormalisation group evolution, the gaugino
masses in the CMSSM scale with constant ratios $M_i^2/g_i^2$, or $M_1
: M_2 : M_3 \sim 1:2:6$. The lightest squark will be the lighter
stop, its \TeV-scale mass mostly depending on $m_{1/2}$.  The
gravitino is similar in mass to the other superpartners and usually
does not affect the LHC phenomenology significantly because it is not
the LSP. A variant of the CMSSM is the non-universal Higgs mass ansatz
(NUHM),
%\index{NUHM|see{supersymmetry!non-universal Higgs mass ansatz}}
%\index{supersymmetry!non-universal Higgs mass ansatz}
which
separates the scalar supersymmetric partner masses from the scalar
Higgs sector.

An alternative breaking mechanism is gauge mediation
(GMSB).
%\index{supersymmetry!gauge mediation}
%\index{GMSB|see{supersymmetry!gauge mediation}} 
If we assume that some kind of messenger sector communicates SUSY 
breaking to the MSSM spectrum, gravity mediation assumes that
this messenger sector is not charged under any of the Standard 
Model charges. In gauge mediation we allow for non-gravitational
interactions between the SUSY breaking sector and the MSSM 
particles~\cite{Dine:1981za,Dimopoulos:1981au}. 
The corresponding messenger particles can have
masses in the \unit{\power{10}{4}}{\GeV} range. Again, the gaugino
masses scale like constant ratios $M_i/g_i^2$. Gaugino masses and
scalar squared masses are generated at the one-loop and two-loop
levels such that eventually they are of similar size. The gravitino
can be the LSP, which for the LHC means that the second-lightest
supersymmetric partner (NLSP) 
can be long-lived before it eventually
decays e.g.\ to a gravitino and a photon or an electroweak gauge
boson.

In such a scenario where the NLSP is linked to the rest of the 
supersymmetric spectrum and the LSP is very weakly coupled, the 
lightest supersymmetric
partner relevant for LHC analyses is different from the actual LSP at cosmological
time scales. 
Now, the NLSP does not have to be electrically
neutral, so the non-gravitational decay chains at the LHC can as well
end with the lightest slepton. The question if we can observe and
measure such late decays at the LHC depends on the lifetime of the
NLSP.

%%%%%%%%%%%%%%%%%%%%%%%%%%%%%%%%%%%
%%%%%%%%%%%%%%%%%%%%%%%%%%%%%%%%
\subsection{Signatures of SUSY}\label{sec:susy:theory:signatures}

While our theoretical understanding of supersymmetry breaking might
make it appealing to search for CMSSM and GMSB signatures right away,
we should instead identify the most generic features of supersymmetric
models and search for those. In the spirit of the pMSSM, we can start
with the straight-forward assumption that the LSP is the lightest
neutralino and that it is responsible for (part of) the dark matter in
the universe.

\begin{figure}[t]
  \begin{center}
    \includegraphics[width=0.65\textwidth]{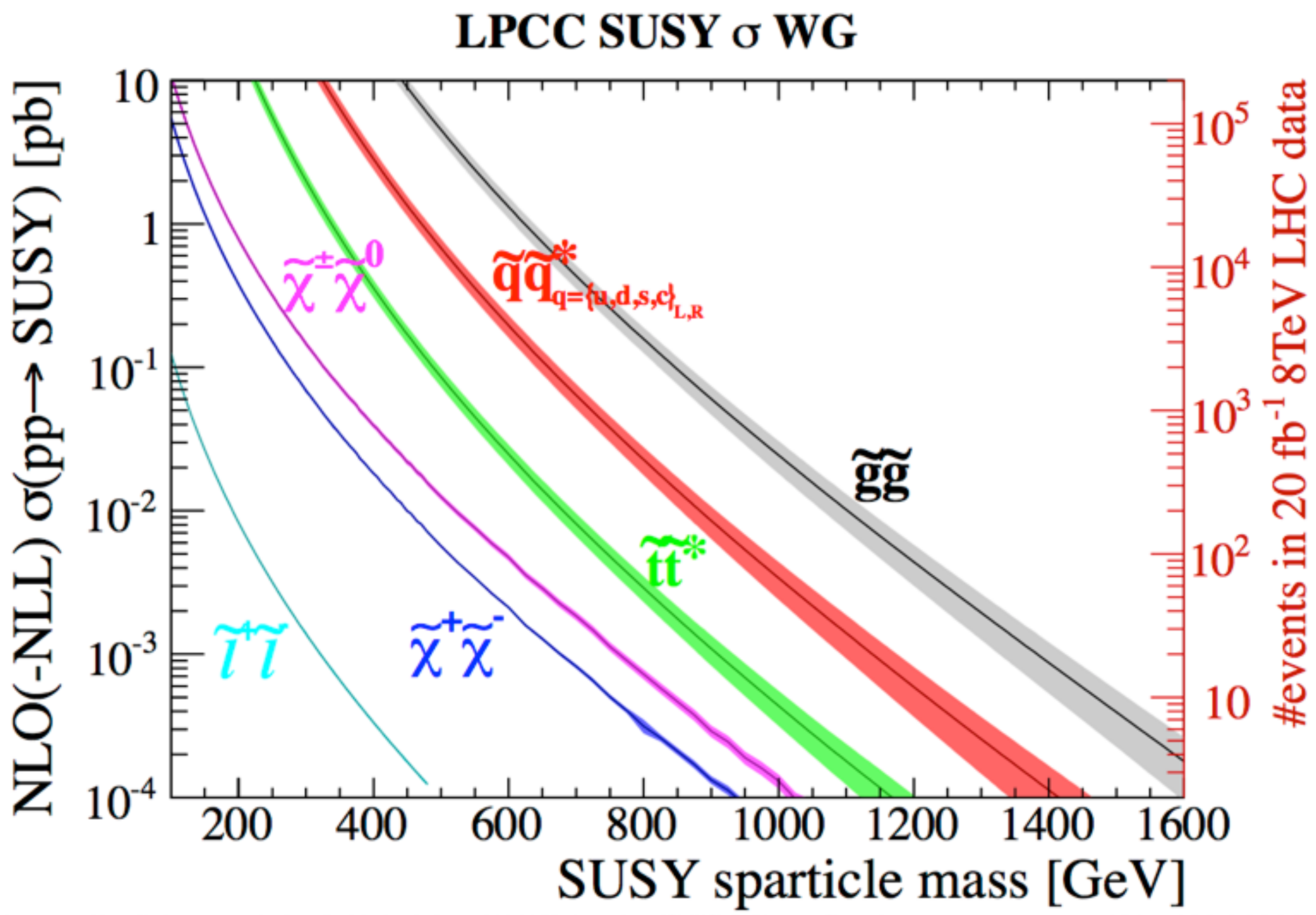} 
  \caption{The cross sections expected for different production modes
    of SUSY particles as a function of their mass. At the same mass
    scale, gluino production dominates over the production of
    first-generation squarks. Third-generation squark production is
    suppressed, since these particles cannot be produced in $t$-channel
    processes. Particles not participating in the strong interaction,
    like gauginos and sleptons, have the smallest production cross section. 
  \textit{(Adapted from Ref.~\cite{Kramer:2012bx}.)}
  \label{fig:SUSY:Expectations:XS}}
  \end{center}
\end{figure}

In that case we expect to see strongly and weakly interacting
supersymmetric states at the LHC. Unless they are too heavy, the
strongly interacting particles with production cross sections
proportional to $\alpS\squared$ will be produced much more frequently
than the weak states with cross sections proportional to $\alpW^2
\sim \alpS\squared/250$.  
Some production cross sections are shown in
\fig{\ref{fig:SUSY:Expectations:XS}} as a function of the mass of the
produced supersymmetric particles.  Because of the Majorana nature of
the $t$-channel gluino exchange, squarks $\tilde{\qq}$  and anti-squarks 
$\tilde{\qq}^*$ can be produced in pairs as 
$\pp \to \tilde{\qq} \tilde{\qq}^*$ and $\pp \to \tilde{\qq}
\tilde{\qq}$. 
Slepton-pair production is further suppressed because
it can only proceed through $s$-channel diagrams. On the other hand,
the LSP cannot be strongly interacting, which means that the directly
produced coloured states first need to shed their colour charge by
radiating jets, thus decaying to the weakly interacting sector,
and then in turn decay to two LSPs.  The resulting LHC signature
consists of jets plus missing energy. A distinct advantage of this
signature is that irreducible background cross sections like
$\Wb+$jets and $\Zb+$jets are not huge and produce much softer jets
compared to the signal. Top-quark pairs can be a dangerous background
and will then have to be suppressed, as described below.

Following the above argument of strongly interacting superpartner
production with a cascade decay to the weak sector and the LSP at the
end, such a decay chain starting from each gluino or squark could read
%
%\begin{linenomath}
\begin{equation}
(\tilde{g} \to ) \; \tilde{\qq} \to \tilde{\chi}_2^0 \to \tilde{\ell} \to
\tilde{\chi}_1^0 \, .
\end{equation}
%\end{linenomath}
%
The corresponding classic kinematic signature for
cascade decays arises from the two leptons: two same-flavour
opposite-sign leptons with an invariant mass below an edge 
given by
%
%\begin{linenomath}
\begin{equation}
0 < m^2_{\ell\ell} 
  < \frac{(m_{\tilde{\chi}^0_2}^2-m_{\tilde{\ell}}^2)
          (m_{\tilde{\ell}}^2-m_{\tilde{\chi}^0_1}^2)}{m_{\tilde{\ell}}^2} \, .
\label{eq:SUSY:edge}
\end{equation}
%\end{linenomath}
%
Typical backgrounds are $\gamma^*/\Zo \to \ell^+ \ell^-$ events as
well as leptonically decaying top-quark pairs. The latter can feature both 
same-flavour and opposite-flavour leptons, which means that they can be
controlled by looking at the difference of the same-flavour and
opposite-flavour events.  The dominant Drell--Yan background for
same-flavour leptons should be well understood experimentally and
theoretically.\index{Drell--Yan processes}  

Changing the production process from light-flavour squarks or sbottoms
to stop pairs leads to a very different decay
signature, for example $\tilde{\tq} \to \tq \tilde{\chi}^0_1$ or
$\tilde{\tq} \to \bq \tilde{\chi}^+_1$. The kinematic edge typical for
cascade decays does not feature in this process, and the final state
can be indistinguishable
from top-quark pair production. 
Hence, in particular for
small stop masses this signal is challenging to extract from the much
larger top-pair background.  Other ways of searching for
supersymmetry, based on electroweak production processes, long-lived
particles or $R$-parity violating decays will be discussed below.

%%%%%%%%%%%%%%%%%%%%%%%%%%%%%%%%
%%%%%%%%%%%%%%%%%%%%%%%%%%%%%%%%%%%
\section{Generic Searches for Supersymmetry}
\label{sec:susy:genericsearches}

Given the large number of free parameters (see \sect{\ref{sec:susy:theory:breaking}}), 
the MSSM in particular and SUSY in general have a plethora of realisations. 
As discussed before, there are various possible signatures
and some of these provide a generic sensitivity to many SUSY models.
In this section we discuss various classes of such signatures and describe a few analysis 
methods used to estimate the remaining background from Standard Model processes after typical event selections.

%%%%%%%%%%%%%%%%%%%%%%%%%%%%%%%%%
%%%%%%%%%%%%%%%%%%%%%%%%%%%%%%%%%%
\subsection{Searches with Jets and \ETmiss}
\label{sec:susy:genericsearches:jets}

In proton-proton collisions, coloured sparticles are dominantly produced via the strong 
interaction---at least if they are not too heavy. 

Although leptons can also be produced (see next \sect{\ref{sec:susy:genericsearches:leptons}}), 
fully hadronic decay chains have a large probability to be realised. 
The LSP at the end of the sparticle decay chain is only weakly interacting and therefore escapes the detector unseen, leaving a momentum imbalance in the transverse plane of the event. This imbalance manifests itself as the missing transverse energy \ETmiss of an event. 
Thus, searches in final states with jets and \ETmiss but without leptons are expected to have 
a good sensitivity to a large fraction of supersymmetric models.

In many cases, SUSY events are expected to produce on average more \ETmiss, or higher multiplicities of energetic jets.
Consequently, searches for new physics are typically selecting events in the tails of distributions of 
%observables like e.g.\ 
\ETmiss and/or other observables, and these quantities have to be precisely understood. 
However, detector effects, misreconstruction or unknown higher-order corrections to 
the production of many additional partons in the final state
are contributing to potentially large systematic uncertainties. To minimise this uncertainty,
most of the searches are using data as much as possible to estimate the SM backgrounds from events 
in phase-space regions which are dominated by SM processes. An example for this approach is Ref. ~\cite{Chatrchyan:2014lfa} which describes the search 
for new physics in final states with many jets and missing 
transverse energy. 
The background to this search has four major contributions:

Firstly, the irreducible background from $\Zb(\to \nu\nu)+$jets events. This background can be 
estimated 
from selected $\Zb(\to l^+l^-)+$jets events, 
by removing the two leptons from the event (as if they would be neutrinos) 
and correcting for lepton efficiencies and acceptance effects as well as for the ratio of the 
branching fractions of $\Zb$ into neutrinos and $Z$ into charged leptons. 
As this ratio is approximately six, this correction
leads to large statistical uncertainty. Alternatively, $\gamma+$jets events can be used by exploiting their
similarity to $\Zb+$jets events at high boson momenta where mass 
effects can be neglected. The statistical uncertainty for this method is significantly smaller,
but the systematic uncertainties from unknown higher-order electroweak and QCD corrections on the cross section 
ratio of $\Zb+$jets and $\gamma+$jets are particularly large for high jet multiplicities.

Secondly, there is a background originating from $\Wb+$jets and $\ttbar$ events 
with leptonically decaying \Wb bosons, i. e.\ $\Wb \to \ell\nu$ with $\ell=e,\mu$. 
The neutrino leads to genuine
\ETmiss, and if the lepton is lost (i.e.\ not reconstructed, not isolated, or out of acceptance), 
the event can be selected. The background can be estimated from a 
$\mu+$jets control sample by reweighting the events according to the 
muon inefficiencies, which are measured from data.

Thirdly, if a $\Wb$ in $\Wb+$jets or $\ttbar$ events decays into a $\tau$ lepton and a neutrino, 
the $\tau$ can decay hadronically and leave a jet-like signature. This contribution can also be 
estimated from a $\mu+$jets control sample by replacing the muon by a jet with a 
momentum expected from a hadronic tau decay. Corrections have to be applied for the 
muon reconstruction and isolation efficiencies, 
as well as for 
the branching ratio of the $\tau$ lepton into hadronic final states.

A final significant
background is QCD multi-jet production with \ETmiss originating from jet energy 
mismeasurements. This background is estimated from an inclusive jet sample.
Each event is, in a first step, rebalanced in the transverse plane, i.e. the jet momenta are adjusted within their uncertainties such that the event has no \ETmiss. 
In a second step, each jet is smeared according to measured jet energy response distributions. 
It is particularly important to include the non-Gaussian tails in the jet energy response, 
as they can lead to 
particularly large values of \ETmiss.

\begin{figure}[t]
  \begin{center}
    \includegraphics[width=0.75\textwidth]{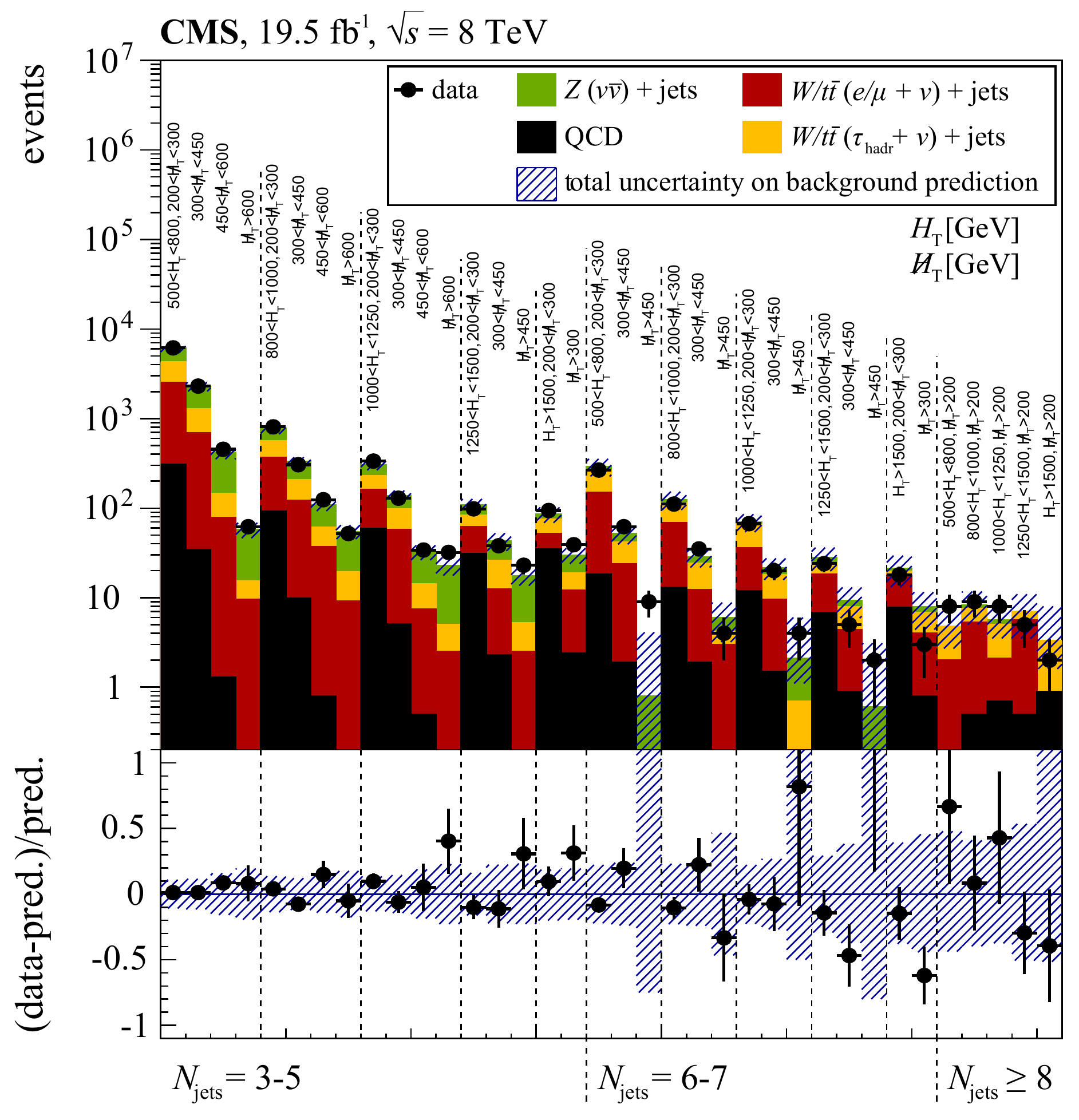}
  \caption{Comparison of background prediction and data for various search regions 
  defined in \HT, \HTmiss, and $N_{\text{jets}}$.
  \textit{(Adapted from Ref.~\cite{Chatrchyan:2014lfa}.)}
  \label{fig:SUSY:CMS_JetsMHT}}
  \end{center}
\end{figure}

Finally, 36 exclusive search regions are defined in $\HT=\sum_{\text{jets}} \pT$, 
$\HTmiss=\vert\sum_{\text{jets}} \vec{p}_{\text{T}}\vert$, 
and $N_{\text{jets}}$. 

Alternative methods for the background estimation are, for example, the usage 
of simulated events, which are then reweighted according to the ratio 
to data in background-dominated control regions.

\begin{figure}[t]
  \begin{center}
    \includegraphics[width=0.95\textwidth]{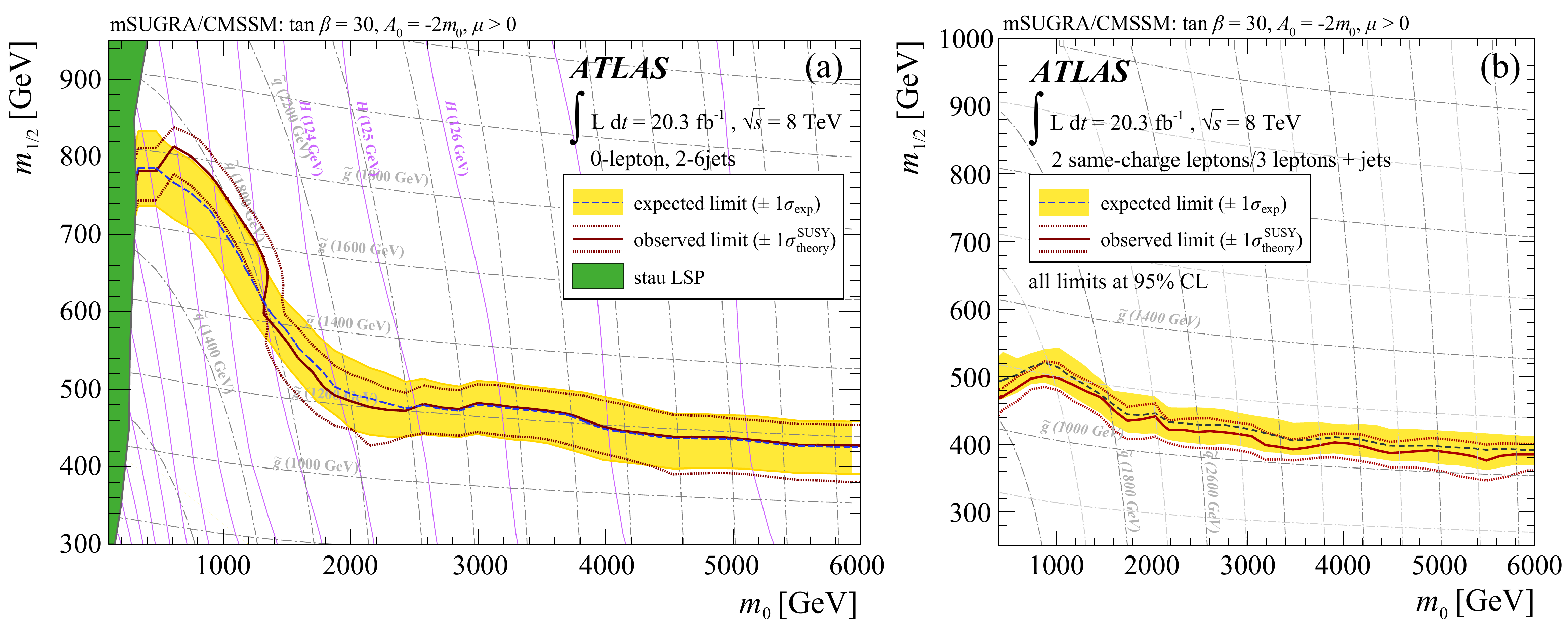}
  \caption{Interpretation of ATLAS searches in final states with jets+\ETmiss or 
  jets+same-sign leptons+\ETmiss in the CMSSM.
  \textit{(Adapted from Refs.~\cite{Aad:2014wea, Aad:2014pda}.)}
    \label{fig:SUSY:ATLAS_CMSSM}
} 
  \end{center}
\end{figure}

A comparison of background prediction and data shows a good agreement 
(see \fig{\ref{fig:SUSY:CMS_JetsMHT}}), and 
since no signal is observed upper limits on sparticle masses are set 
in various supersymmetric models.
If the results are interpreted in the CMSSM,
\index{SUSY!constrained MSSM}
as shown 
in \fig{\ref{fig:SUSY:ATLAS_CMSSM}}(a), 
the fully hadronic searches are sensitive to squark masses up to \unit{1.3}{\TeV} 
and gluino masses up to \unit{1.8}{\TeV}.
Mass limits in less constrained models can be significantly weaker, 
as will be discussed in \sect{\ref{sec:susy:status:SMS}}.

Many other analyses with different observables and analysis techniques 
have been carried out in the fully hadronic final state.
One often used observable is $m_\text{eff} = \HT + \ETmiss$, which is an estimator 
for the total energy scale of the event~\cite{Aad:2014wea}.
Other variables are motivated by the process $\tilde{q}\tilde{q}\to q\tilde{\chi}^0_1q\tilde{\chi}^0_1$ 
for which the two jets originating from the emitted quarks are possibly enclosing a 
small azimuthal angle ($<\pi$), while for QCD dijet events the two leading jets are mostly 
back-to-back in the transverse plane.
Moreover, the momenta of the jets are carrying some information on the mass difference 
between the initially produced SUSY particles and the final-state LSPs.
Therefore, the jet kinematics can be used to construct observables which are having 
a high potential to separate background from signal, and which are further sensitive to the 
scale of a given scenario of new physics.
In the case of longer decay chains, the jets are clustered in two ``hemispheres'', and for each of them the four momenta of the jets are added to one ``pseudo-jet''.
The resulting two pseudo-jets are then used for further kinematic investigations.
Three of the variables which have been used in the fully hadronic channel 
are $\alpha_\text{T}$~\cite{Randall:2008rw, Chatrchyan:2013lya}, the 
razor~\cite{Chatrchyan:2012uea}, and 
the ``stransverse mass''
$m_{\text{T}2}$~\cite{Lester:1999tx,Barr:2003rg,Chatrchyan:2012jx,Khachatryan:2015vra}, 
which is further explained in \sect{\ref{sec:susy:EWthirdsearches}}.
All the different approaches show a comparable sensitivity to new physics.

%%%%%%%%%%%%%%%%%%%%%%%%%%%%%%%%%
%%%%%%%%%%%%%%%%%%%%%%%%%%%%%%%%%%
\subsection{Final States with Leptons and \ETmiss}
\label{sec:susy:genericsearches:leptons}

Leptons can be produced in supersymmetric decay 
chains either if sleptons are 
present in the sparticle decay chain or if heavier neutralinos or charginos occur. 
In the latter case, $\Wb$ and $\Zb$ bosons can be produced, which can decay further into leptons. 
If the initially produced supersymmetric particle is a squark or a gluon, 
at least one jet per decay chain is produced, and thus another typical final-state signature 
for many supersymmetric models consists of jets+isolated leptons+\ETmiss, where the isolation observable is calculated from further energy deposits in a cone around the lepton. 
We distinguish here between leptons of the first two generations (electrons and muons) 
and of the third generation ($\tau$ leptons) since the reconstruction of 
hadronically decaying $\tau$ leptons is more complex. 
Different lepton multiplicities and charge combinations are sensitive to different models: 
Single-lepton searches are quite generic, but often suffer from large backgrounds from 
Standard Model processes~\cite{Aad:2012ms, Chatrchyan:2012ola}. 
Final states with three or more leptons are particularly sensitive to the direct production 
of electroweak superpartners, as will be discussed later. 
Final states with two leptons can be separated into different classes according to 
the lepton charges and flavours: same sign or opposite sign with same flavour or opposite flavour. 
In particular the searches in final states with same-sign leptons provide 
good sensitivity to many models, because in signal events the two leptons with 
the same charge can be produced in different decay chains, e. g.\ $\tilde{\qq}\tilde{\qq}\to \qbar\qbar\tilde{\chi}^\pm_1\tilde{\chi}^\pm_1\to \qbar\qbar l^\pm\nu l^\pm\nu \tilde{\chi}^0_1\tilde{\chi}^0_1$, while 
Standard Model same-sign processes are very rare and mostly originate from events 
where a non-prompt lepton, e. g.\ from a heavy-flavour jet, accidentally
fulfills the isolation requirement~\cite{Aad:2014pda, Chatrchyan:2013fea}. 
This background is usually estimated by selecting leptons with a looser isolation criterion 
and reweighting the events with a tight-to-loose ratio that has been measured in 
a background-enriched control region.

So far, no excess beyond the Standard Model expectation has been established at the LHC for 
searches in final states with leptons, jets and \ETmiss, and the results are interpreted in various models 
of supersymmetry. 
Figure~\ref{fig:SUSY:ATLAS_CMSSM}(b) shows the interpretation in 
the CMSSM.
\index{SUSY!constrained MSSM}
While the jets+\ETmiss searches 
provide the best sensitivity at small values of $m_0$ (see Figure~\ref{fig:SUSY:ATLAS_CMSSM}(a)), the sensitivity of searches with same-sign 
leptons are almost independent of $m_0$, and both searches exclude at large values of $m_0$ comparable regions in the $m_0$--$m_{1/2}$ plane.

\begin{figure}[t]
  \begin{center}
    \includegraphics[width=0.95\textwidth]{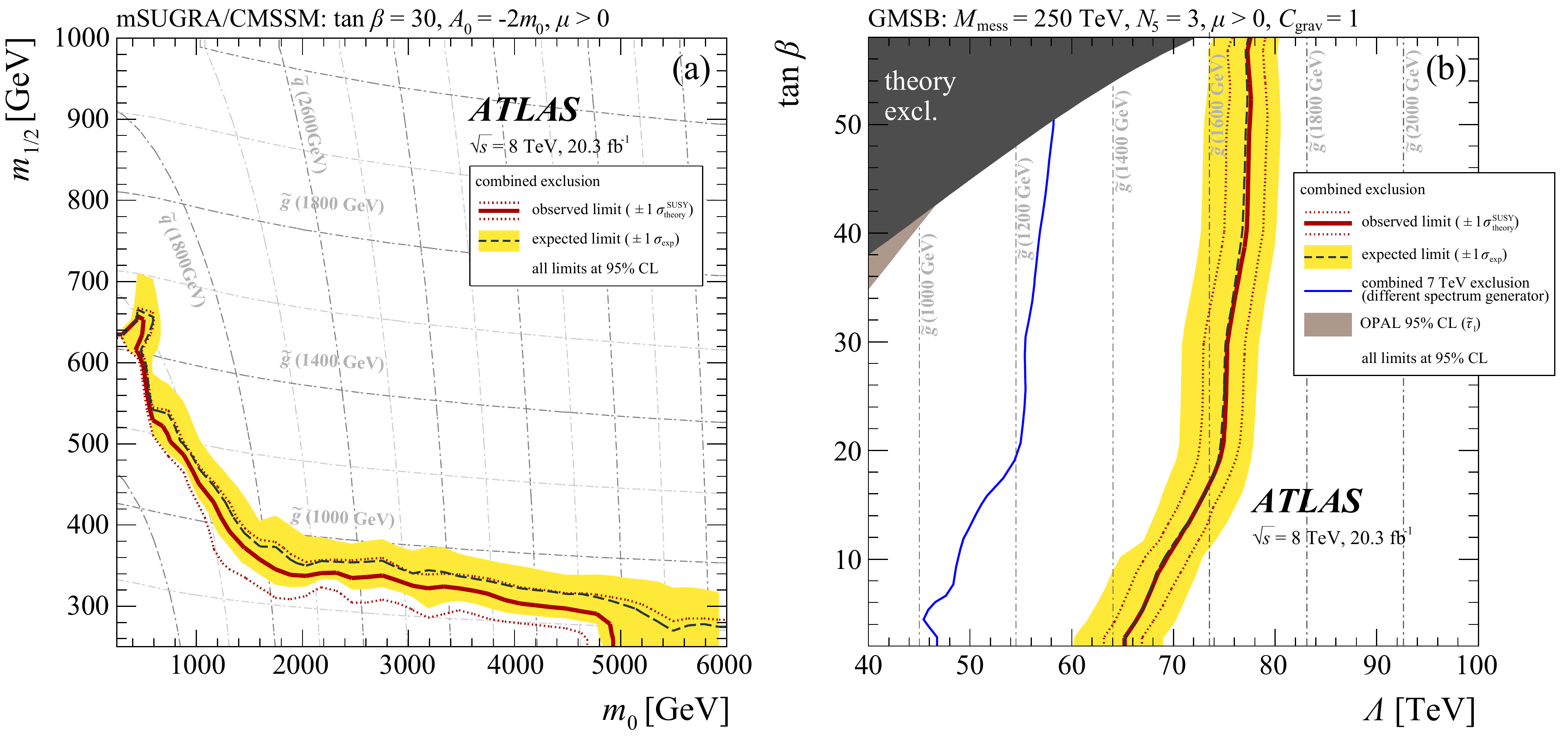}
  \caption{(a) CMSSM interpretation of the ATLAS search with at least one $\tau$ 
  lepton, \ETmiss and jets. The search performs best for small $m_0$, i. e.\ small 
  $\tilde{\tau}$ masses. 
  (b) Interpretation of the same analysis in the GMSB; the limits are shown in the plane 
  defined by the parameter $\tan\beta$ and the effective SUSY breaking scale $\Lambda$. 
  \textit{(Adapted from Ref.~\cite{Aad:2014mra}.)}
  \label{fig:SUSY:tau_CMSSM_GMSB}
  } 
  \end{center}
\end{figure}

In many supersymmetric models the partner of the $\tau$ lepton, 
the stau ($\tilde{\tau}$)
is the lightest slepton. Light staus that are almost mass-degenerate with the 
LSP are very interesting from a cosmological point of view as will be further 
explained in \sect{\ref{sec:susy:status:Fits}}. Thus, if staus
are light, it is expected that they occur frequently in decay chains, and that 
consequently numerous
$\tau$ leptons are produced. Signatures in which 
the $\tau$ lepton decays further into a light lepton and two neutrinos are covered 
by the searches with electrons and/or muons as described above. 
However, in $\sim 65\%$ of the cases, the $\tau$ decays into an odd number of 
charged hadrons accompanied by neutral hadrons. 
If the mass splitting between the stau and the LSP is small, the 
$\tau$ leptons can have rather low transverse momenta, and not all might be reconstructed. 
Therefore, searches with only one or with
more $\tau$ leptons 
in final states with additional jets from the decay of the initially produced squarks or gluons and \ETmiss
are motivated~\cite{Chatrchyan:2013dsa, ATLAS:2012ht, Aad:2014mra}. 
The dominant background for the search with one $\tau$ lepton are 
$\Wb+$jets 
events and, to a smaller extent, QCD multi-jet events with one 
jet being misreconstructed as a hadronically decaying $\tau$.
For the final state with two $\tau$ leptons, the dominant backgrounds are 
again $\Wb+$jets events with one real $\tau$ and one misidentified jet, 
as well as $\ttbar$ events with two real $\tau$ leptons. No evidence for new physics has 
been observed by ATLAS and CMS so far, 
and limits on supersymmetric parameters have been set.
Two interpretations within the
CMSSM and within
the GMSB model
%\index{GMSB|see{gauge mediated symmetry breaking}}
%\index{gauge mediated symmetry breaking|see{supersymmetry}}
%\index{supersymmetry!gauge mediation}
are displayed in \fig{\ref{fig:SUSY:tau_CMSSM_GMSB}}. Especially for the 
CMSSM interpretation it can be seen that the best sensitivity is observed
for models where $m_0$ and thus the mass of the stau is small
and close to the mass of the LSP.

Furthermore, a very generic search for anomalous production of three or four leptons 
in various flavour combinations and kinematic search regions, e.g.\ with low 
or high accompanying jet activity, has been performed~\cite{Chatrchyan:2014aea}.
Once more, no significant excess beyond the expected statistical fluctuations has been observed. 

%%%%%%%%%%%%%%%%%%%%%%%%%%%%%%%%
%%%%%%%%%%%%%%%%%%%%%%%%%%%%%%%%%%%
\subsection{Final States with \ETmiss and Photons}
\label{sec:susy:genericsearches:photons}

\begin{figure}[t]
  \begin{center}
    \includegraphics[width=0.95\textwidth]{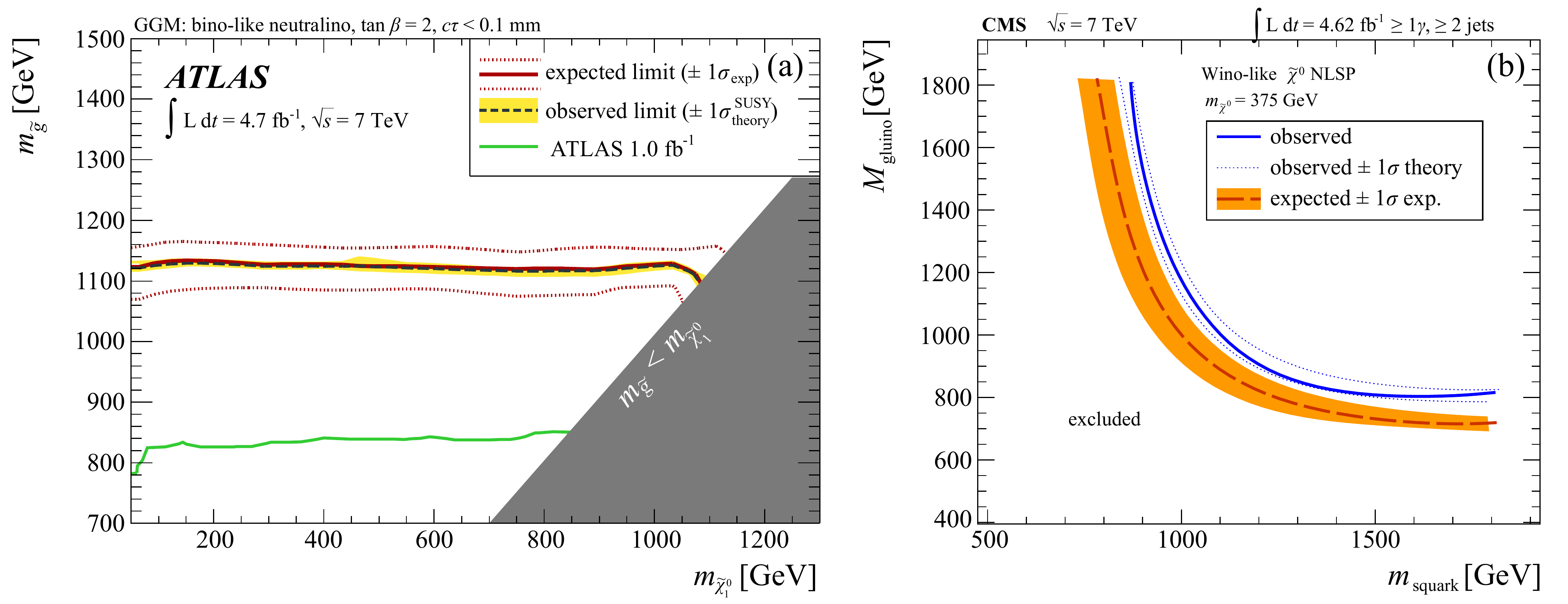}
  \caption{Interpretation of the searches with photons, jets and \ETmiss in the GGM.
  (a) Exclusion limit from a search with two photons in the plane spanned by the NLSP 
  mass and the gluino mass for a bino-like NLSP. 
  (b) Exclusion limit 
  from a search with at least one photon in the plane spanned by the squark mass and the gluino mass for a fixed wino-like NLSP mass.
  \textit{(Adapted from Refs.~\cite{Aad:2012zza, Chatrchyan:2012bba}.)}
  \label{fig:SUSY:GGM}
  } 
  \end{center}
\end{figure}

In the searches discussed above, it was assumed that the LSP is the lightest neutralino. 
However, in GMSB models the gravitino is  
the LSP~\cite{Dimopoulos:1996yq}. As---due to the weakness 
of gravitation---the 
couplings of the gravitino to other particles are heavily suppressed, all decay chains are reaching 
the next-to-lightest supersymmetric particle (NLSP), 
%\index{next to lightest supersymmetric particle}
%\index{NLSP|see{next-to-lightest supersymmetric particle}}
\index{SUSY!LSP}
which then decays further into the gravitino and a SM particle. 
The type of this SM particle 
is related to the type of the NLSP: If the NLSP is ``bino-like'', 
the SM particle will most likely be a photon, while for ``wino-like'' NLSPs it will in most cases be 
a \Zb boson. This fact motivates searches for supersymmetry in final states with 
one or more photons in association with jets and \ETmiss. 
Important backgrounds for such searches are originating from photon+jet processes 
with jet energy mismeasurement leading to artificial \ETmiss, from 
QCD multi-jet production where one of the jets is identified as a photon, 
and lastly from $\Wb(\to e\nu)+$jets production with the electron being misreconstructed as 
a photon. As no ATLAS or CMS search shows a significant excess beyond the 
Standard Model expectation, mass limits have been set in various models. 
Figure~\ref{fig:SUSY:GGM} shows two interpretations within general 
gauge mediation models (GGM)
%\index{GGM|see{general gauge mediation}}
%\index{general gauge mediation|see{supersymmetry}}
%\index{supersymmetry!general gauge mediation}
models~\cite{Meade:2008wd}, 
with a bino-like or wino-like NLSP~\cite{Aad:2012zza, Chatrchyan:2012bba}
(GGM is a generalisation of GMSB models).

%%%%%%%%%%%%%%%%%%%%%%%%%%%%%%%%%%%%%%%%%%%%%%%%%%%%
\subsection{Simplified Models: Virtues and Challenges}
\label{sec:susy:status:SMS}

In Sections \ref{sec:susy:genericsearches:jets} to \ref{sec:susy:genericsearches:photons}, various results from SUSY searches have 
been interpreted in the context of so called ``full models''.
\index{SUSY!full models}
These are characterised by a relatively small number of parameters that fully specify 
the sparticle mass spectrum and thereby fix the production cross sections and 
branching ratios of every sparticle.
In such models events can be in general realised in many different ways, 
e. g.\ by different primarily produced sparticles and by different combinations of the two 
decay chains.
Although this provides realistic signatures for any
assumed SUSY model with any set of specific parameters, 
the results are difficult to generalise to other models or event topologies.
Another limitation is that in a given full model, the masses of different sparticles 
are often related in a way that can not be generalised. In the CMSSM,
\index{SUSY!constrained MSSM} 
for example, 
there is the well known $\sim 1:2:6$ relation for 
$M_1:M_2:M_3$.

\begin{figure}[t]
  \begin{center}
    \includegraphics[width=0.95\textwidth]{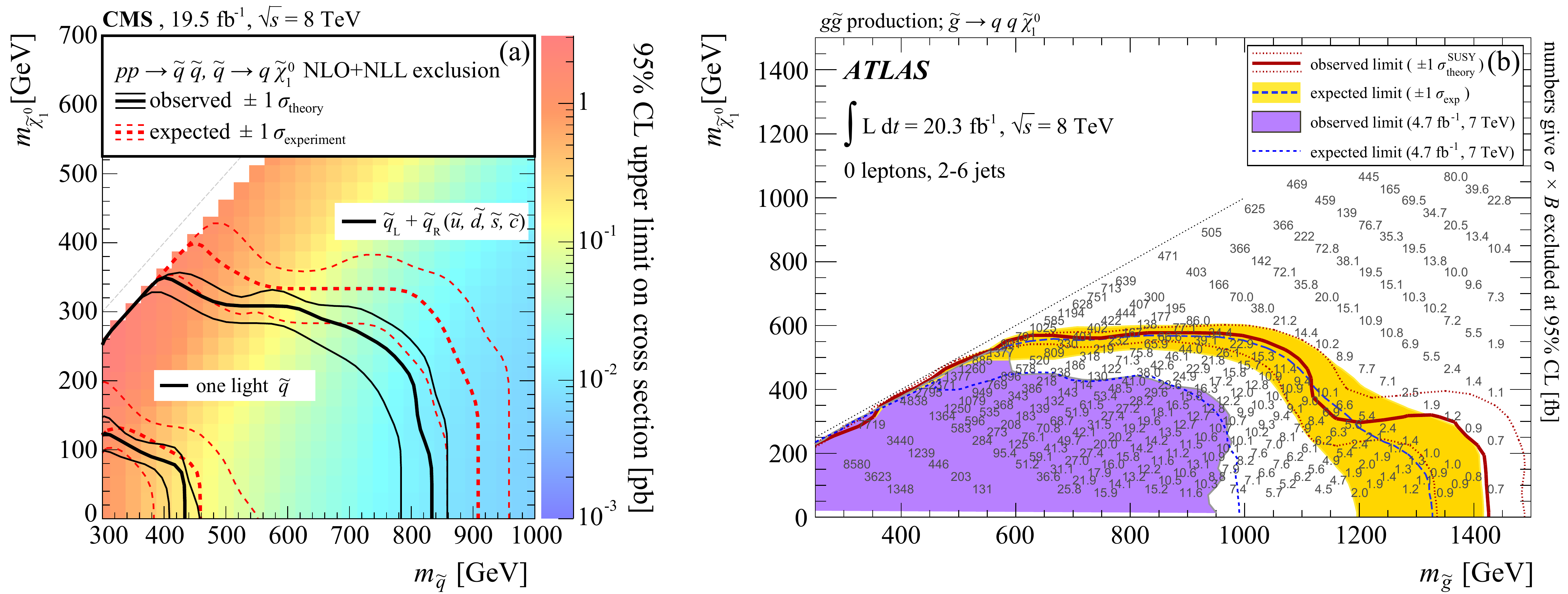}
  \caption{(a) Interpretation of a CMS search with jets+\ETmiss 
  in a simplified model with 100\% branching ratio 
  of $\pp\to\tilde{\qq}\tilde{\qq}\to \qq\tilde{\chi}_1^0 q\tilde{\chi}_1^0$, with 
  only one or four squark flavours 
  kinematically accessible, shown in the $m(\tilde{q})$--$m(\tilde{\chi}_1^0)$ plane.
  (b) A similar search by ATLAS interpreted in a simplified model with the decay 
  topology $\pp\to\tilde{g}\tilde{g}\to \qq\qbar\qq\qbar\tilde{\chi}^0_1\tilde{\chi}^0_1$ shown in 
  the $m(\tilde{g})$--$m(\tilde{\chi}_1^0)$ plane. 
  \textit{(Adapted from Refs.~\cite{Chatrchyan:2014lfa,Aad:2014wea}.)}
  \label{fig:SUSY:CMS_T2qq}
  }
  \end{center}
\end{figure}

An alternative approach is the concept of simplified model 
searches (SMS)
%\index{SMS|see{simplified models}}
%\index{simplified models|see{supersymmetry}}
\index{SUSY!simplified models}
~\cite{Chatrchyan:2013sza}: 
The basic assumption here is that only one particular event topology is realised, 
e.g.\ $\pp\to\tilde{\qq}\tilde{\qq}\to \qq\tilde{\chi}_1^0 q\tilde{\chi}_1^0$.
The only free parameters are then the masses of the two involved sparticles, 
namely $m_{\tilde{\qq}}$ and $m_{\tilde{\chi}_1^0}$, which make these ``models'' rather simple 
and easy to interpret.
The production cross section is fixed by the squark mass~\cite{Kramer:2012bx}, which 
potentially depends on the number of kinematically accessible squark flavours, 
as can be seen in \fig{\ref{fig:SUSY:CMS_T2qq}}.

With the data from Run~1, searches with jets and \ETmiss are sensitive to 
gluino masses up to \unit{1.3}{\TeV} for events of the type
$\tilde{\glue}\tilde{\glue}\to \qq\qbar\qq\qbar\tilde{\chi}^0_1\tilde{\chi}^0_1$, 
and squark masses up to \unit{880}{\GeV} for $\tilde{\qq}\tilde{\qq}\to \qq\qq\tilde{\chi}^0_1\tilde{\chi}^0_1$ (see \fig{\ref{fig:SUSY:CMS_T2qq}}).

Obviously, the assumption of a branching ratio of 100\% in a
particular final state leads to overly optimistic results if other
final states with signatures that are difficult to find are realised for a
significant fraction of the events. Therefore, the exclusion limit for a full model can only be achieved by the
interpretation of the SMS 
result within this full model. An example for
such an analysis is presented in \sect{\ref{sec:susy:status:Fits}}.

%%%%%%%%%%%%%%%%%%%%%%%%%%%%%%
%%%%%%%%%%%%%%%%%%%%%%%%%%%%%%%%%%%%%
\section[SUSY Searches for Electroweak and Third-Generation Production]{The
  Rest of the Spectrum: SUSY Searches for Electroweak and Third-Generation Production}
\label{sec:susy:EWthirdsearches}

In the previous section, searches for SUSY were discussed that aim at
the most abundant signatures in generic implementations of SUSY. These
generic signatures are also featured in the most common highly
constrained models such as the CMSSM, where the mass scales of all
sfermions are coupled to each other and all gauginos are coupled to
each other. Thus, excluding high first-generation squark (and gluino)
masses means that in models like the CMSSM also light sleptons 
and third-generation squarks and gauginos are indirectly excluded. 
In this section, searches are covered for which the simple assumption of
maximally constrained SUSY shall not be a motivation.

The generic searches are motivated by the relatively high cross
section of gluino production and first-generation and
second-generation squark production, as shown in
\fig{\ref{fig:SUSY:Expectations:XS}}. However, as shown e.g.\ in
\fig{\ref{fig:SUSY:ATLAS_CMSSM}}, the limits obtained in this type of
search in highly constrained SUSY models such as the CMSSM correspond
to SUSY mass scales of gluinos and squarks of more than
$M_{\tilde{\glue}}=\unit{1.2}{\TeV}$ and
$M_{\tilde{\qq}}=\unit{1.8}{\TeV}$, respectively. If all SUSY masses
were that heavy, SUSY would cease to help the hierarchy problem, as
discussed in \sect{\ref{fig:SUSY:ATLAS_CMSSM}}.

Therefore, in order to retain the feature of solving the hierarchy
problem, it is theoretically attractive to think of more general SUSY
models in which the mass scale of the sparticles that only interact
weakly (the sleptons and gauginos) and the mass scale of the
third-generation squarks are decoupled from the scale of the first-generation
and second-generation squarks, and from that of the gluinos.
It is easily possible to realise such a model in the framework of the MSSM or in
even more general models. On the other hand, there is no obvious choice for a benchmark
model for such a scenario, defined at the GUT scale with only a very small number of
parameters. This is in contrast to the generic searches. 
Thus, the results of the specific searches for SUSY in electroweak
production processes or in the production of third-generation squarks
ask for a more model-independent parametrisation, to allow for the
later interpretation of the search results in any specific SUSY
model. This is mostly realised in the form of the so-called simplified
model searches already introduced in the previous section. While this
section covers the searches and the translation of their results into
SMS limits, the application of these limits is discussed further in
\sect{\ref{sec:susy:status:Fits}}.

%%%%%%%%%%%%%%%%%%%%%%%%%%%%%%%%
%%%%%%%%%%%%%%%%%%%%%%%%%%%%%%%%%%%
\subsection{Electroweak Production of SUSY Particles}
\label{sec:susy:EWthirdsearches:EW}

\begin{figure}[t]
  \begin{center}
    \includegraphics[width=0.7\textwidth]{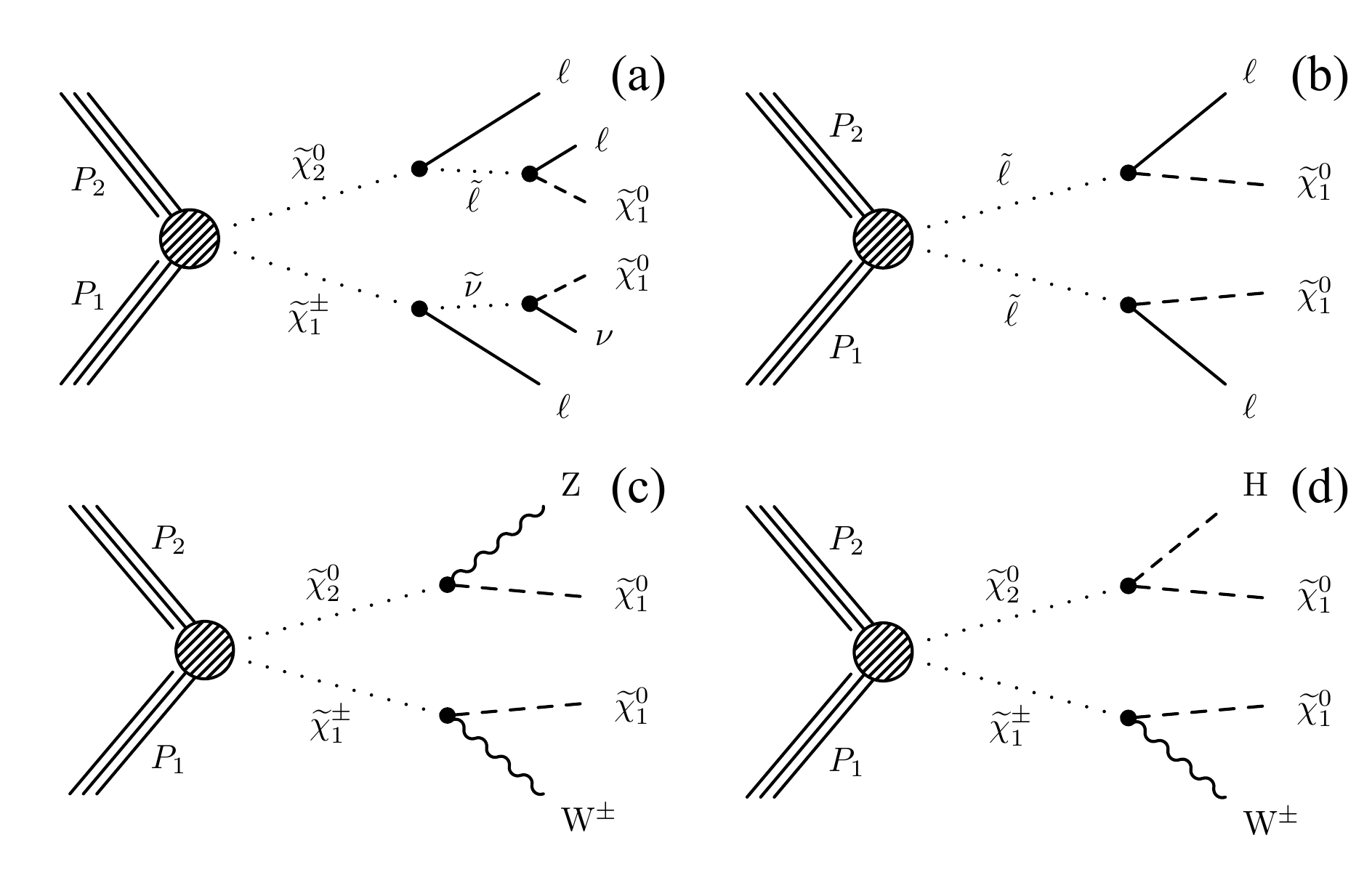}
  \caption{Example diagrams for the production of gauginos or sleptons
    without the production of strongly interacting sparticles. These
    production processes involve electroweak interactions.
   \label{fig:SUSY:EW_feynman_diagrams}
   }
  \end{center}
\end{figure}

Searches for SUSY events with electroweak interactions at the
production vertex can be classified in two different ways: first, by
the type of SUSY particles occurring at the production vertex, and
second, by the visible final-state particles. Examples for production
and decay processes are shown in
\fig{\ref{fig:SUSY:EW_feynman_diagrams}}.

It can be seen that some of the fundamental features of SUSY events
already discussed in \sect{\ref{sec:susy:genericsearches}} are
retained: Leptons in the final state allow events to be distinguished
from the QCD background, and in $R$-parity conserving models missing
transverse energy \ETmiss is produced by the escaping LSPs, here
assumed to be the lightest neutralinos $\tilde{\chi}^0_1$.  However, there are
also significant differences with respect to generic searches for SUSY
events involving the strong interaction at the production vertex:
There, the high production cross section of up to ${\mathcal
  O}(\unit{10}{\pico\barn})$ (depending on the mass scale, see
%fig:SUSY:Expectations:XSf
\fig{\ref{fig:SUSY:Expectations:XS}}) made possible by the strong
interaction allows rather heavy SUSY objects to be produced at the
\unit{8}{\TeV} LHC Run~1. However, in electroweak production processes
for SUSY particles not interacting strongly, there is a reduction in
the production cross section at the same mass scale of more than a
factor of 100. Thus, at the same centre-of-mass energy, the accessible
mass scales for the direct production of sleptons and gauginos is
significantly lower than for first-generation squarks and
gluinos. Consequently, also the transverse momenta of the final-state
particles and the expected \ETmiss are significantly lower than for
strong-production processes of heavy squarks and gluinos---a fact that can
limit the selective power of the trigger and the offline selection.

While gauginos couple directly to the quarks in the protons via gauge
interactions, the sleptons need an intermediate SM gauge boson to
couple to the partons in the beam protons. This explains why the
direct production of sleptons again is suppressed with respect to the
gaugino production by about a factor of 100.

As shown in \fig{\ref{fig:SUSY:EW_feynman_diagrams}}, different
production modes can yield very similar final states. Leptonic decays
of \Wb and \Zb bosons are employed, which means that the first three
diagrams yield near-identical final-state particle compositions of two
or more leptons and missing energy. Depending on the intermediate
particles, the expected kinematics may vary. On the other hand, the usable decays
of Higgs bosons, e.g.\ into \bq quarks, $\tau$ leptons or two photons,
yield final states with two hadronic heavy quark jets or two photons.
Therefore, a strategy is employed where the searches are classified by
final-state particles, and each search can then be interpreted in
simplified models describing different production mechanisms, which
can be optimally constrained by an individual final state. In the
following, out of a very large number of available
searches~\cite{Aad:2014nua,Aad:2014vma,Aad:2014iza,Aad:2014yka,Chatrchyan:2013fea,Khachatryan:2014qwa,Khachatryan:2014mma},
the following examples are discussed in detail: the production of
2-lepton and 3-lepton final states in conjunction with \ETmiss, where
the leptons can be electrons or muons; the production of $\tau$-lepton
final states with \ETmiss; and the specific search for SUSY decays
with electroweak production of SUSY particles with a Higgs boson in the decay
chain.

The first example discussed here stems from ATLAS and CMS searches for
2 or 3 leptons and \ETmiss, where the leptons can come in same-sign or
opposite-sign configurations~\cite{Aad:2014vma,Khachatryan:2014qwa}.
It can be beneficial to also search for same-sign lepton
configurations in cases where one out of three leptons is outside of
the detector acceptance in production processes as shown in
\fig{\ref{fig:SUSY:EW_feynman_diagrams}}(a).  The dominant background
to these searches are diboson events ($\WW$, $\Wpm\Zb$ and
$\ZZ$), $\Wpm+$jets and $\Zb+$jets events, and $\ttbar$ production with a
small contamination of leptons from secondary decays, e.g.\ from heavy
quarks.

\begin{figure}[t]
  \begin{center}
    \includegraphics[width=0.95\textwidth]{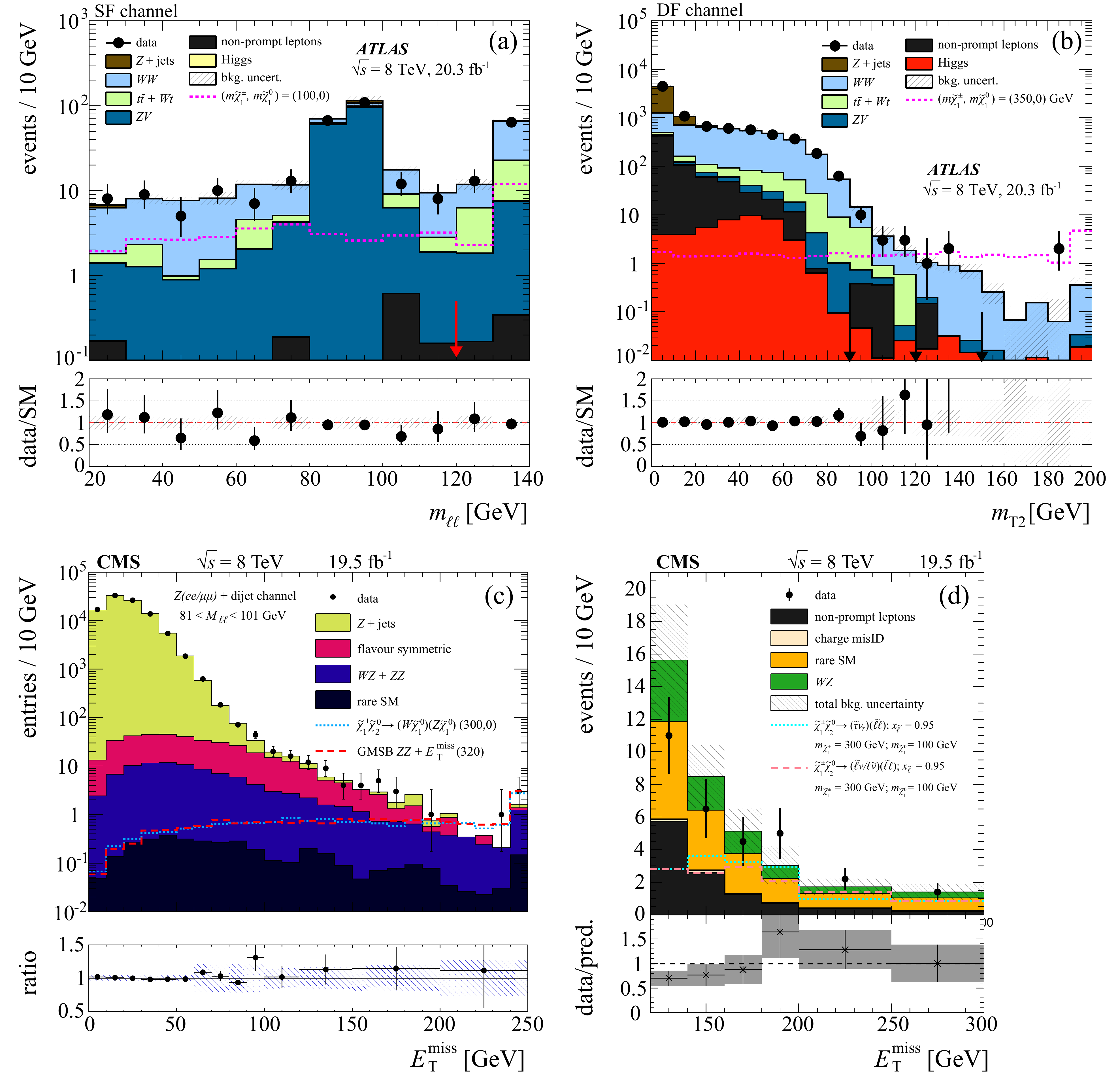}
  \caption{Selected observables from the searches for the production
    of gauginos and sleptons via the electroweak interaction from 
    ATLAS (a,b) and CMS (c,d). 
    See the text for a
    discussion of the observables, signal properties and backgrounds.
    \textit{(Adapted from Refs.~\cite{Aad:2014vma,Khachatryan:2014qwa}).}
  \label{fig:SUSY:EW2l3l:Observables}
        }
  \end{center}
\end{figure}

Examples for discriminating variables employed in these searches
are given in \fig{\ref{fig:SUSY:EW2l3l:Observables}}. An example for a
signal region aiming at
SUSY decays involving two \Wpm bosons is
shown in \fig{\ref{fig:SUSY:EW2l3l:Observables}}(a) from the ATLAS experiment. 
The plot shows the distribution of the invariant mass of two
opposite-sign same-flavour (SF) leptons
$m_{\ell\ell}$, where $\ell=e,\mu$. 

An example of another signal region is shown in \fig{\ref{fig:SUSY:EW2l3l:Observables}}(b). 
It exploits the ``stransverse'' mass
\mTtwo for two leptons of different flavours (DF)~\cite{Lester:1999tx,Barr:2003rg}, defined as
\[
\mTtwo = \min_{\qTvec}\left[\max\left(\mT(\pTell{1},\qTvec),\mT(\pTell{2},\pTmiss-\qTvec)\right)\right]\, ,
\]
where $\pTell{1}$ and $\pTell{2}$ are the transverse momenta of the
two leptons, and $\qTvec$ is a transverse vector that minimises the
larger of the two transverse masses $\mT$.  The latter is defined by
\[
\mT(\pTvec,\qTvec) = \sqrt{2(\pT\qT-\pTvec\cdot\qTvec)}\, .
\]
For SM \ttbar and \WW events, in which two \Wpm bosons decay
leptonically and $\pTmiss$ originates from the two neutrinos, the
\mTtwo distribution has an upper end-point at the \Wpm mass.  For
signal events, the undetected LSP contributes to $\pTmiss$, and the
\mTtwo end-point is correlated to the mass difference between the
slepton or chargino and the lightest neutralino.  For large values of
this difference, the \mTtwo distribution for signal events extends
significantly beyond the distributions obtained from \ttbar and \WW
events. This is clearly visible in
\fig{\ref{fig:SUSY:EW2l3l:Observables}}(b), where the background
drops strongly on the logarithmic scale at about $\mTtwo\approx
\MW$, while the signal stays flat.

Other typical observables besides the leptonic invariant mass,
\mTtwo and \mT, are $\HT$ (see
\sect{\ref{sec:susy:genericsearches:jets}}), \ETmiss and
$m_{\text{eff}}=\HT+\ETmiss$. Examples for \ETmiss are given in
\fig{\ref{fig:SUSY:EW2l3l:Observables}}(c,d), for the signal regions
aimed at
decays via \Wpm and \Zb (c) and for decays of the gauginos
via staus (d). The latter is favourable in many models motivated by
the stau co-annihilation mechanism of reducing the predicted amount of
dark matter in the universe down to the measured values (see
\sect{\ref{sec:susy:status:Fits}}).\index{dark matter}  There, the stau is the NLSP and
thus is kinematically favoured in the decay of heavier SUSY particles.
It can be seen that the SM backgrounds drop strongly for growing
\ETmiss, and only SM processes with true \ETmiss from e.g.\
$\Zb\to\nu\bar{\nu}$ decays remain at large values. On the other hand,
due to the energy carried by the LSP, SUSY processes on average show
much higher \ETmiss values compared to the background.
Thus, high signal-to-background ratios can be reached at very high \ETmiss.

Neither ATLAS nor CMS observed a significant SUSY signal in
any of these signal regions in Run~1.  
Thus, the results of the
searches are converted into limits on SMS decay chains, as already
discussed in \sect{\ref{sec:susy:status:SMS}}. Since different signal
regions can be sensitive to different signals depending on the
kinematical configuration, expected limits from all signal regions are
derived for all model points and decay chains. The observed limit for
a given decay chain and model point is then calculated only for the
signal region with the strongest expected limit.

\begin{figure}[t]
  \begin{center}
    \includegraphics[width=0.95\textwidth]{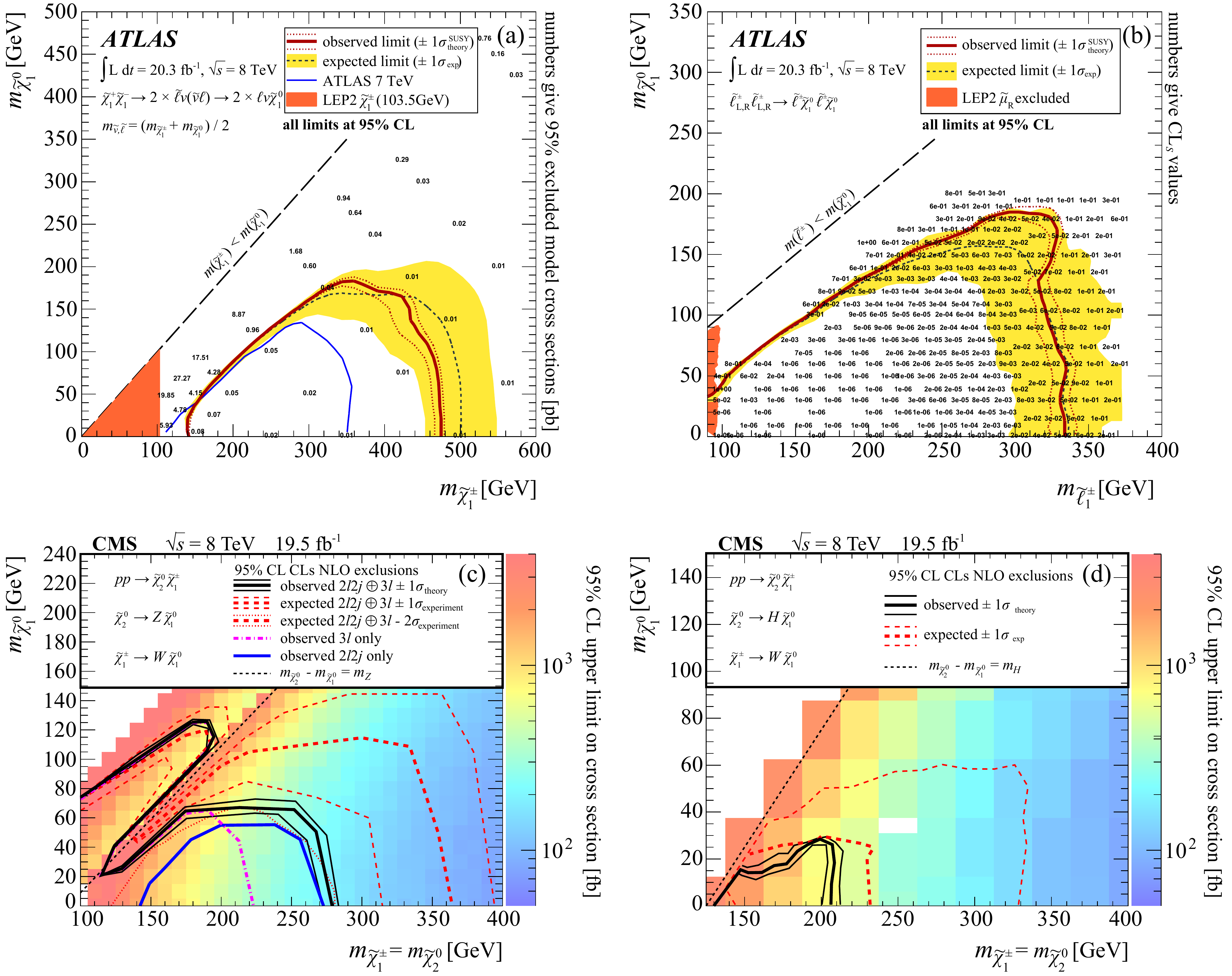}
  \caption{The resulting limits on selected SMS decay chains from
    ATLAS (a,b) and CMS (c,d). The excluded
    production cross section of the decay chain given in the figures
    is plotted against the masses of the physical SUSY particles
    appearing in the decay chain. The expected exclusion lines in the
    plots are given by the assumption of the full electroweak coupling
    at the production and of 100\% branching fractions into the
    studied final state. Individual kinematical assumptions on the
    particles involved in the decay are also given in the plots.
    \textit{(Adapted from Refs.~\cite{Aad:2014vma,Khachatryan:2014qwa}.)}  
    \label{fig:SUSY:EW2l3l:Limits}
}
  \end{center}
\end{figure}

Examples for the limits derived for the searches described above are
given in \fig{\ref{fig:SUSY:EW2l3l:Limits}}. The top row shows
examples from the ATLAS experiment, the bottom row from the CMS
experiment. The considered SMS decay chains are $\tilde{\chi}_1^+\tilde{\chi}_1^-\to
\tilde{\ell}\nu/\tilde{\nu}\ell\to 2\ell\nu\tilde{\chi}_1^0$ with
$m_{\tilde{\ell},\tilde{\nu}}=(m_{\tilde{\chi}_1^{\pm}}-m_{\tilde{\chi}_1^0})/2$
(\fig{\ref{fig:SUSY:EW2l3l:Limits}}(a)),
$\tilde{\ell}^+\tilde{\ell}^-\to2\ell^{\pm}\tilde{\chi}_1^0$ (b),
$\tilde{\chi}_2^0\tilde{\chi}_1^{\pm}\to Z\tilde{\chi}_1^0W\tilde{\chi}_1^0$ (c), and
$\tilde{\chi}_2^0\tilde{\chi}_1^{\pm}\to h\tilde{\chi}_1^0W\tilde{\chi}_1^0$ (d).  The physical masses
of the SUSY particles involved are given on the axes of the plots,
while the numbers (ATLAS) respectively color scale (CMS) 
give the cross
section of the exclusive production of the decay chain under investigation
that can be excluded at the 95\% \CL.  
If only two SUSY masses are involved, this limit on the cross section
of individual decay chains can be interpreted directly in every SUSY
model which involves the decay chain, as long as there is no strong
dependence on the production process. It is possible that---depending
on the nature of the gauginos in a given physical model---the relative
strengths of the predicted $s$-channel and $t$-channel production
processes do vary,
which could influence
the selection efficiency and thus the limit on the
cross section. Therefore, in some cases $t$-channel production is
removed from the simulated process by assuming the particle in the $t$-channel, e.g.\ a squark in
the SMS implementation, to be very heavy. However, influences from
differences in the production process for the same decay chains are
typically weak.

A more challenging situation arises for decay chains with more than
two different SUSY particles that are not degenerate in mass.  
One example is the decay chain 
$\tilde{\chi}_1^+\tilde{\chi}_1^-\to
\tilde{\ell}\nu/\tilde{\nu}\ell\to 2\ell\nu\tilde{\chi}_1^0$ decay chain in
\fig{\ref{fig:SUSY:EW2l3l:Limits}}(a)  with the
explicit assumption of
$m_{\tilde{\ell},\tilde{\nu}}=(m_{\tilde{\chi}_1^{\pm}}-m_{\tilde{\chi}_1^0})/2$. Also
this latter condition has an effect on the acceptance and efficiency of the
search. However, this assumption is not generically true in all
physical models to which the SMS limit shall be applied. Thus, the
correct interpretation of such limits might involve additional studies
of changes in the detector acceptance between SMS implementation and
physical model.

A general feature of the limits is that, for specific mass relations,
the momenta of the visible decay products become small. For the decay
chains that proceed via the emission of SM fermions with negligible
mass (see \fig{\ref{fig:SUSY:EW2l3l:Limits}}(a,b)), this is the case
for $m_{\tilde{\chi}_1^{\pm}}=m_{\tilde{\chi}^0_1}$ or
$m_{\tilde{\ell}}=m_{\tilde{\chi}^0_1}$, respectively. For the cases of
\fig{\ref{fig:SUSY:EW2l3l:Limits}}(c,d), where heavy SM gauge bosons
are emitted from the decay chain, the kinematical condition in which
the sensitivity vanishes changes to
$m_{\tilde{\chi}_1^{\pm}}-m_{\tilde{\chi}^0_1}=m_{\Vb,\Hb}$. In these cases, the trigger
and selection efficiencies degrade considerably, leading to a very
weak exclusion. This is one of the two reasons
why---unfortunately---no general statement in the form of ``gauginos
(or sleptons) are excluded up to a mass of'' can be derived from the
LHC search results.

\begin{figure}[t]
  \begin{center}
    \includegraphics[width=0.95\textwidth]{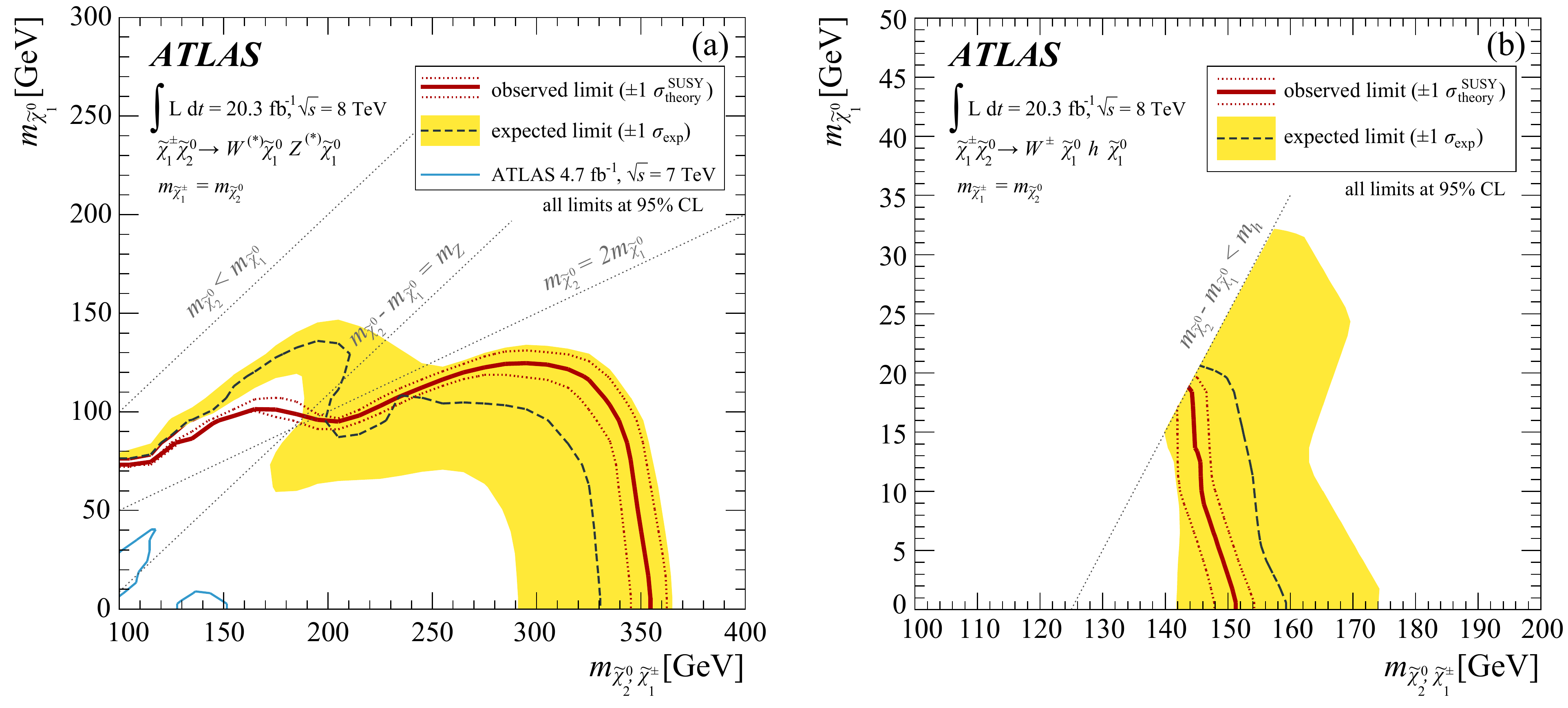}
    \caption{Interpretation of ATLAS searches in final states with
      three leptons ($e$, $\mu$ or $\tau$) for direct chargino and
      neutralino production. As expected, the limits on the sparticle
      masses for an assumed simplified model with $\tilde{\chi}^0_2
      \to \Zb + \tilde{\chi}^0_1$ (a) are much stronger than those for the
      case of decay chains involving the SM-like Higgs boson $h$:
      $\tilde{\chi}^0_2 \to h + \tilde{\chi}^0_1$ (b).
      \textit{(Adapted from Ref.~\cite{Aad:2014nua}.)}
  \label{fig:SUSY:ATLAS_NeuCha}
      }
  \end{center}
\end{figure}

The other reason for the lack of universal limits on individual SUSY
production modes can be exemplified as follows:
Figure~\ref{fig:SUSY:ATLAS_NeuCha} shows the interpretation of
leptonic searches with three leptons ($e$, $\mu$ or
$\tau$)~\cite{Aad:2014nua}. Assuming an exclusive branching ratio of
$\tilde{\chi}^0_2 \to \Zb + \tilde{\chi}^0_1$
(\fig{\ref{fig:SUSY:ATLAS_NeuCha}}(a)) is neglecting a possibly large
branching ratio including a SM-like Higgs boson $h$ in 
the form of $\tilde{\chi}^0_2 \to h + \tilde{\chi}^0_1$
(\fig{\ref{fig:SUSY:ATLAS_NeuCha}}(b)).  Both can yield the same final
state, however with strongly different SM branching fractions to the
visible final-state particles and with different yields of $\tau$
leptons versus light leptons.  In a physical model, the real limit
would have to be derived for a combination of the two decay channels given above.
Since the limits derived for the two channels differ significantly due
to the much weaker branching fraction of the Higgs boson into light
leptons, the limit has to be evaluated for each model point depending
on its individual prediction for the branching fractions.

Many other interesting search channels 
are covered along the lines described above by the ATLAS and CMS experiments. Examples are the
search for electroweak SUSY production with $\tau$ leptons
without accompanying jets~\cite{Aad:2014yka,Khachatryan:2014qwa} or
more specific searches with Higgs bosons in the decay
chain~\cite{Chatrchyan:2013mya}.

\begin{figure}[t]
  \begin{center}
    \includegraphics[width=0.65\textwidth]{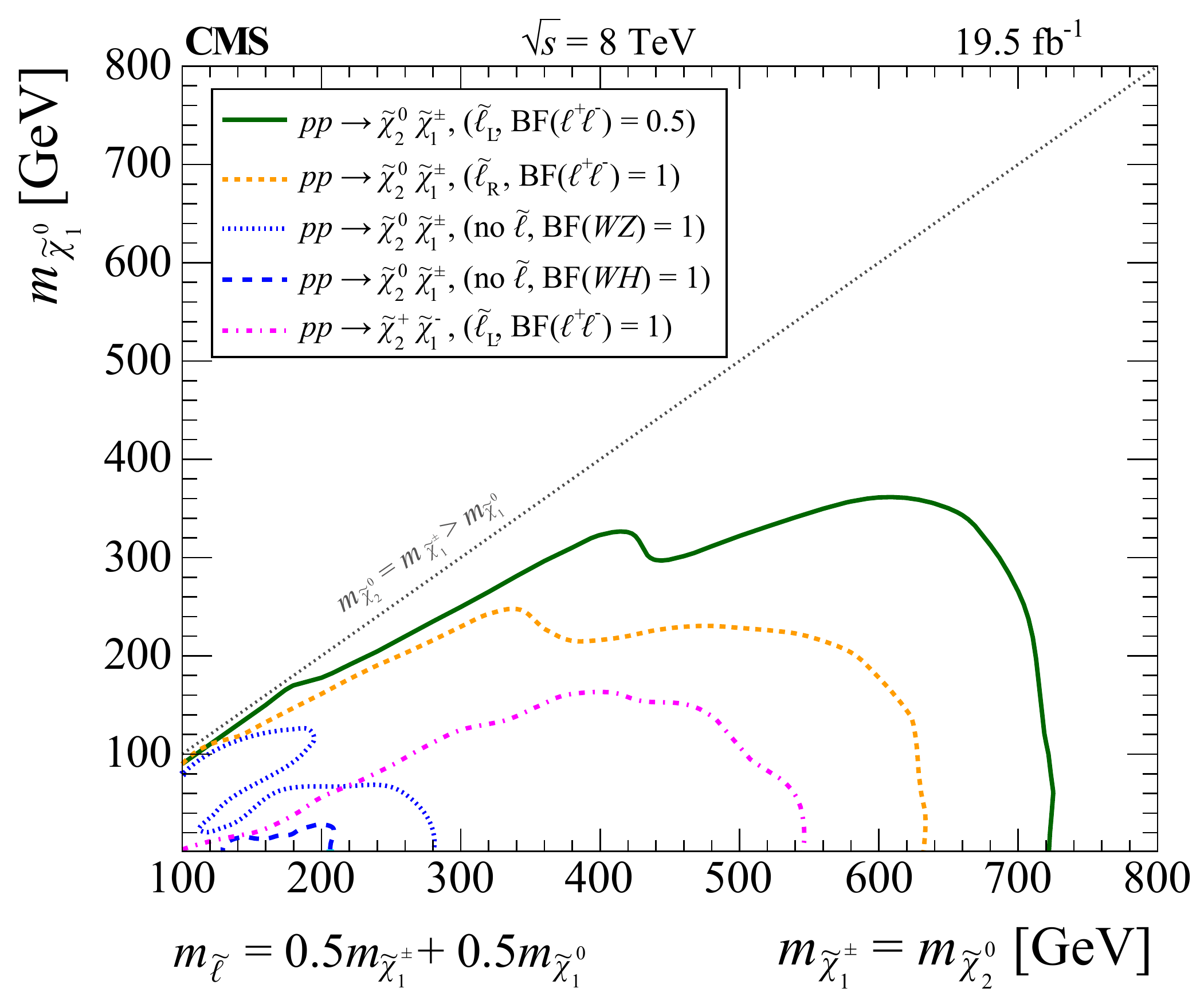}
  \caption{Overview plot of a selection of published limits on gaugino
    production from the CMS collaboration. All limits are given in the
    form of SMS limits on individual production processes.
    Similar results are available from the ATLAS collaboration.
    \textit{(Adapted from Ref.~\cite{Khachatryan:2014qwa}.)}
  \label{fig:SUSY:EWoverview}
    }
  \end{center}
\end{figure}

%herehere
A summary of CMS searches for electroweak production of gauginos is
given in
\fig{\ref{fig:SUSY:EWoverview}}~\cite{Khachatryan:2014qwa,Khachatryan:2014mma}. Similar
results are available from
ATLAS~\cite{Aad:2014nua,Aad:2014vma,Aad:2014yka}. Depending on which
particles are involved, different kinematical bounds exist. The
strongest limits in terms of the accessible mass range are obtained
for $\tilde{\chi}_1^{\pm}\tilde{\chi}_2^0\to
\tilde{\ell}\nu/\tilde{\ell}\ell\to 3\ell\nu2\tilde{\chi}_1^0$ in
specific 3-lepton searches~\cite{Aad:2014nua,Khachatryan:2014qwa} that
show a very striking signature of 3 light leptons with high transverse
momenta and large \ETmiss. For both ATLAS and CMS, masses of up to
$m_{\tilde{\chi}_1^{\pm}}=m_{\tilde{\chi}_2^0}\approx\unit{700}{\GeV}$
can be excluded at the 95\% \CL. However, for small values of
$m_{\tilde{\chi}_1^{\pm}}-m_{\tilde{\chi}_1^0}$, the visible momenta
of the leptons disappear.  Thus, even for an assumption of an
exclusive branching fraction in the given decay mode and full
electroweak coupling strength at the vertex,
$m_{\tilde{\chi}_1^{\pm}}\approx m_{\tilde{\chi}_2^0}\approx
\unit{200}{\GeV}$ is still allowed. In many SUSY breaking scenarios,
however, some fine-tuning of the parameters is needed in order to achieve
$m_{\tilde{\chi}_1^{\pm}}\approx m_{\tilde{\chi}_1^0}$.

%%%%%%%%%%%%%%%%%%%%%%%%%%%%%%%%%%%%%%
%%%%%%%%%%%%%%%%%%%%%%%%%%%%%
\subsection{Searches for SUSY Particles of the Third Generation}
\label{sec:susy:EWthirdsearches:ThirdGen}

As already motivated in \sect{\ref{sec:susy:theory}}, there is a
strong interest in the third-generation squarks. The main reason for
this is twofold: 
First, third-generation squarks play a key role in determining the
mass of the SM-like Higgs boson $h$ in a SUSY model. If their mixing
is large, i.e.\ if large mass splittings between the different mass
eigenstates $\tilde{\tq}_{1}$ and $\tilde{\tq}_{2}$ (or between
$\tilde{\bq}_{1}$ and $\tilde{\bq}_{2}$) occur, then $m_{h}\approx
\unit{126}{\GeV}$ can be achieved without excessively unnatural mass
hierarchies between the SUSY particles and their SM partners.
Furthermore, the large mass splitting yields a situation where
$\tilde{\tq}_{1}$ and $\tilde{\bq}_{1}$ are the lightest predicted
squarks.  Second, the production cross section of third-generation
squarks at the LHC is about one order of magnitude lower than 
the corresponding cross sections of the SUSY partners of light quarks
at the same mass (see \fig{\ref{fig:SUSY:Expectations:XS}}), despite
sharing 
the same couplings. This is a consequence of the negligible
heavy-quark content of the proton.  Therefore, $\tilde{\tq}_{1,2}$ and
$\tilde{\bq}_{1,2}$ can only be produced in $s$-channel processes via
a gluon, and not in the $s$-channel via a quark or in $t$-channel
processes. While the discovery of $\tilde{\tq}_{1,2}$ and
$\tilde{\bq}_{1,2}$ thus would be of highest importance in the context
of the connection between SUSY and the Higgs sector, their
lower cross section makes them more difficult to pin down than the more
abundantly produced first-generation squarks.

\begin{figure}[t]
  \begin{center}
    \includegraphics[width=0.95\textwidth]{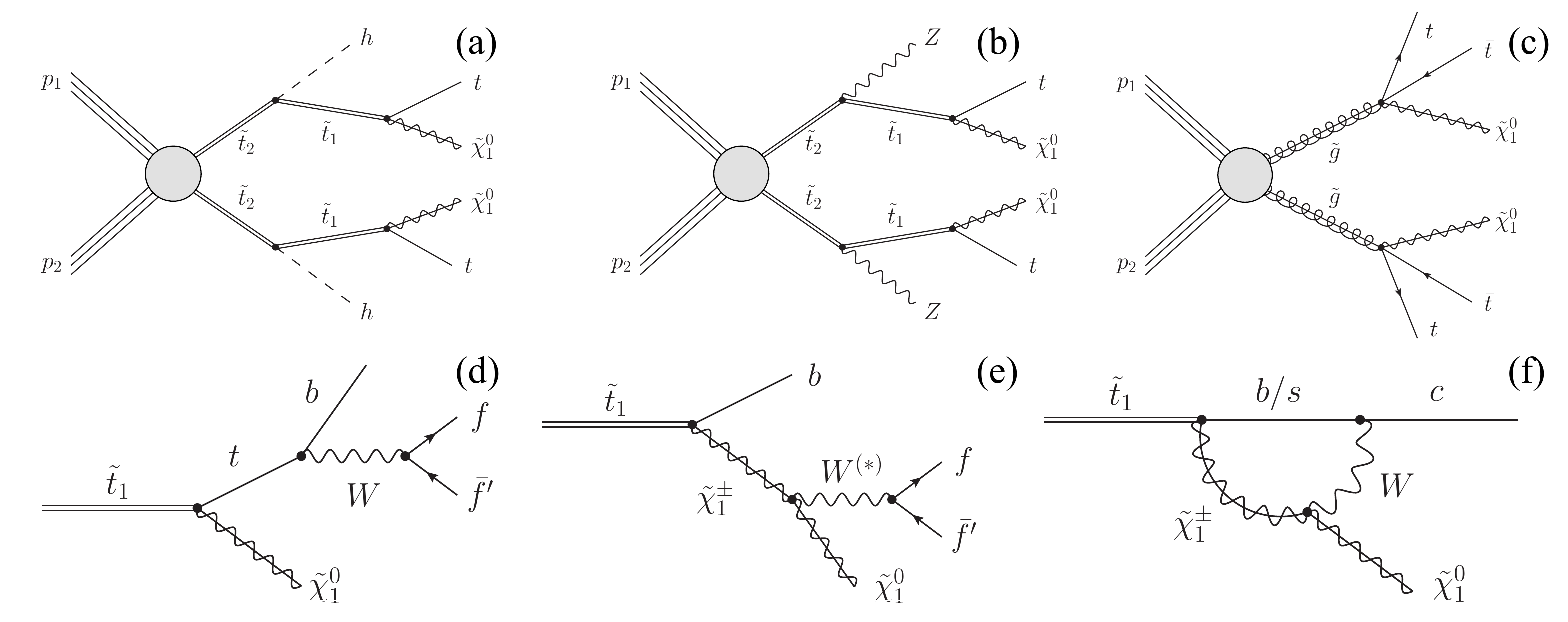}
  \caption{Example diagrams for the production and decay of top quarks
    in conjunction with the production of the lighter or heavier stops or gluinos.  
    \label{fig:SUSY:stop_feynman_diagrams}
    }
  \end{center}
\end{figure}

Figure~\ref{fig:SUSY:stop_feynman_diagrams} shows the typical
production and decay processes of stops. They can either be
produced directly as the heavier mass eigenstate with a decay via a
Higgs boson (a) and via $\Zb$ bosons (b), or via the decay of gluinos
(an example for a three-body-decay of a gluino with a virtual stop is
given in (c)). Further, if the lighter mass eigenstate is produced,
its
decay could proceed via neutralinos (d) or charginos
(e,f). Therefore, a complex collection of different final states can
be searched for, the realisation of which depends both on the couplings of the model and on
the kinematical configuration.

\begin{figure}[t]
  \begin{center}
    \includegraphics[width=0.8\textwidth]{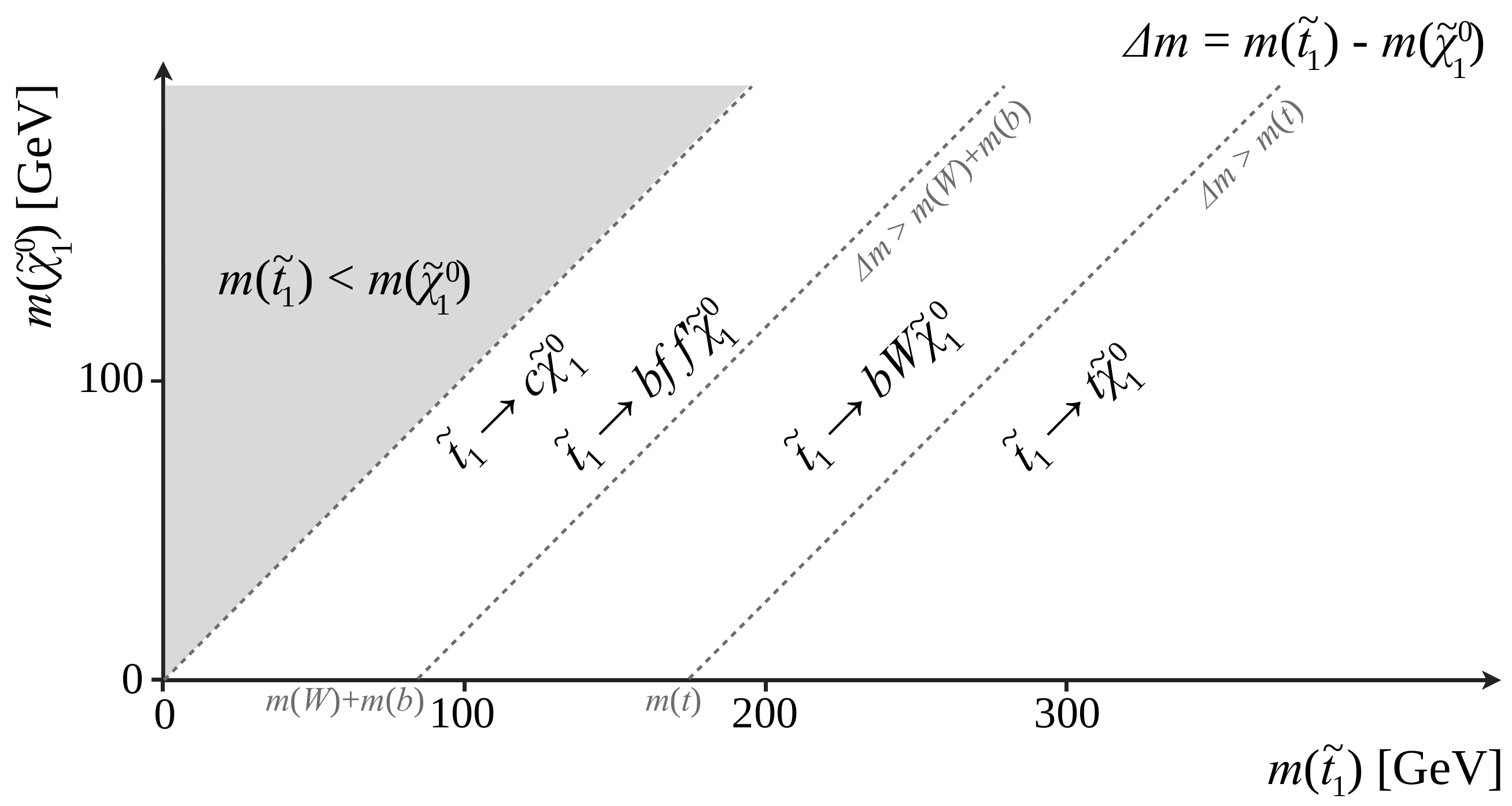}
  \caption{A schematic illustration of the different kinematical areas
    which determine the selection topology of stop
    searches.
    \textit{(Adapted from Ref.~\cite{Aad:2014kra}.)}
  \label{fig:SUSY:thirdGen:Kinematics}
    }
  \end{center}
\end{figure}

These kinematical configurations are shown in
\fig{\ref{fig:SUSY:thirdGen:Kinematics}} in the
$m_{\tilde{\tq}_1}$--$m_{\tilde{\chi}_1^0}$ plane. 
For large $m_{\tilde{\tq}_1}$, the direct two-body decay
$\tilde{\tq}_1\to \tq\tilde{\chi}_1^0$ is open. For very large
$m_{\tilde{\tq}}$, the top quark from the decay is boosted strongly in
the transverse direction, which might require specific reconstruction
techniques, as outlined below. As soon as
$m_{\tilde{\tq}_1}-m_{\tilde{\chi}_1^0}=\Mt$ is reached, the direct two-body
decay is closed and the sensitivity drops to zero. For smaller
$m_{\tilde{\tq}_1}$, the decay via a (virtual) chargino into the same
final state can be used: $\tilde{\tq}_1\to \bq\tilde{\chi}_1^{\pm}\to
\bq\Wpm\tilde{\chi}_1^1$. For $m_{\tilde{\tq}_1}-m_{\tilde{\chi}_1^0}=\Mb+\MWpm$, also
this channel closes and the sensitivity vanishes again. Then, the
decay via virtual $\Wb\to \ensuremath{ff'}$ or the decay
$\tilde{\tq}_1\to \cq\tilde{\chi}_1^0$ remains. Finally, the region
$m_{\tilde{\tq}_1}<m_{\tilde{\chi}_1^0}$ is uninteresting from a cosmological
point of view.

\begin{figure}[t]
  \begin{center}
    \includegraphics[width=0.95\textwidth]{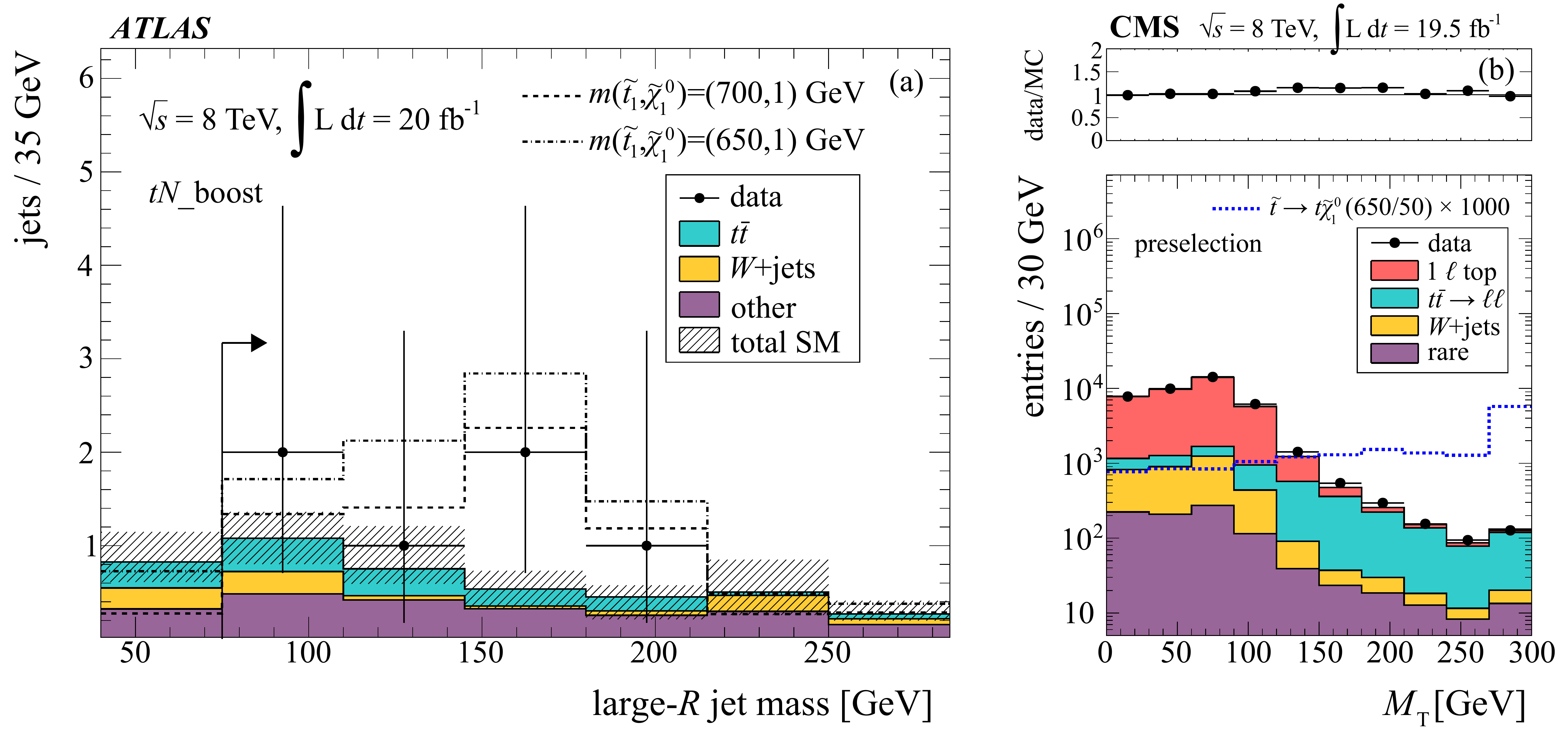}
  \caption{Illustrations of typical selection variables for searches
    for $\tilde{\tq}_1$ decays. (a) The ATLAS
    search for heavy stops yielding fat jets
    shown in the form of the jet mass of the fat jet. 
    (b) An example from CMS with a
    search for intermediate-mass stops
    shown in the form of the transverse-mass distribution.
    \textit{(Adapted from Refs.~\cite{Aad:2014kra,Chatrchyan:2013xna})}
  \label{fig:SUSY:thirdGen:Observables}
    }
  \end{center}
\end{figure}

Out of a very large set of available
searches~\cite{Aad:2013ija,Aad:2014qaa,Aad:2014mha,Aad:2014bva,Aad:2014kva,Aad:2014lra,Aad:2014kra,Aad:2014nra,Chatrchyan:2013xna,Chatrchyan:2013mya},
a few examples shall be explained in the
following. Figure~\ref{fig:SUSY:thirdGen:Observables} shows examples
from the event selections of such searches. For the ATLAS experiment,
an example from a search for heavy stops is
shown~\cite{Aad:2014kra}. For large $m_{\tilde{\tq}_1}\gg
\Mt+m_{\tilde{\chi}^0_1}$, the top quark from the decay and the hadronic decay products of the $\tq\to
\bq\Wb\to \bq\ensuremath{ff'}$ decay are strongly boosted, and the top
quark cannot easily be reconstructed from the individual jets of its 
decay products. 
In this case, the search for so-called ``fat jets'' or
``large-$R$ jets''
%\index{fat jets|see{large-$R$ jets}}
%\index{large-$R$ jets}
is employed. First, an anti-\ktalgo jet with a very wide radius parameter
$R=1.0$, $\pT>\unit{150}{\GeV}$ and $|\eta|<2.0$ is
reconstructed~\cite{Butterworth:2008iy,Aad:2013gja}. Within this jet,
jet trimming~\cite{Krohn:2009th}
%\index{jet trimming} 
is applied with
an anti-\ktalgo jet finder and $R=0.3$. From all objects within the
large-$R$ jets that carry a fraction of more than 0.05 of the
transverse momentum of the large-$R$ jet, an invariant mass is
reconstructed. The distribution of this reconstructed mass is shown in
\fig{\ref{fig:SUSY:thirdGen:Observables}}(a) after preselection
cuts. Flavour tagging is then applied to the sub-jets, which should
contain a $\bq$-quark jet. It can be seen that for signal events, the
jet-mass distribution shows a very broad peak-like structure around
the top-quark mass. The remaining backgrounds are \ttbar and single
top quark production, \ttbar production in association with a vector
boson, $\Zb+$jets, and diboson
production. The data show a slight, but insignificant excess over the
background.

An example is a search from the CMS experiment, which is both
sensitive to $\tilde{\tq}_1\to t\tilde{\chi}_1^0$ and to non-resonant
$\tilde{\tq}_1\to \bq\Wb\tilde{\chi}_1^0$ at intermediate values of
$m_{\tilde{\tq}_1}$~\cite{Chatrchyan:2013xna}.  There, the variable
\mT already introduced in the previous section is used as a
discriminator against events where the real missing transverse energy
is exclusively stemming from a $\Wb\to\ell\nu$ decay, such as
semileptonic \ttbar events. No excess over the background is observed,
and the variable is used as an input to a multivariate selection, from
which limits are derived.

\begin{figure}[t]
  \begin{center}
    \includegraphics[width=0.85\textwidth]{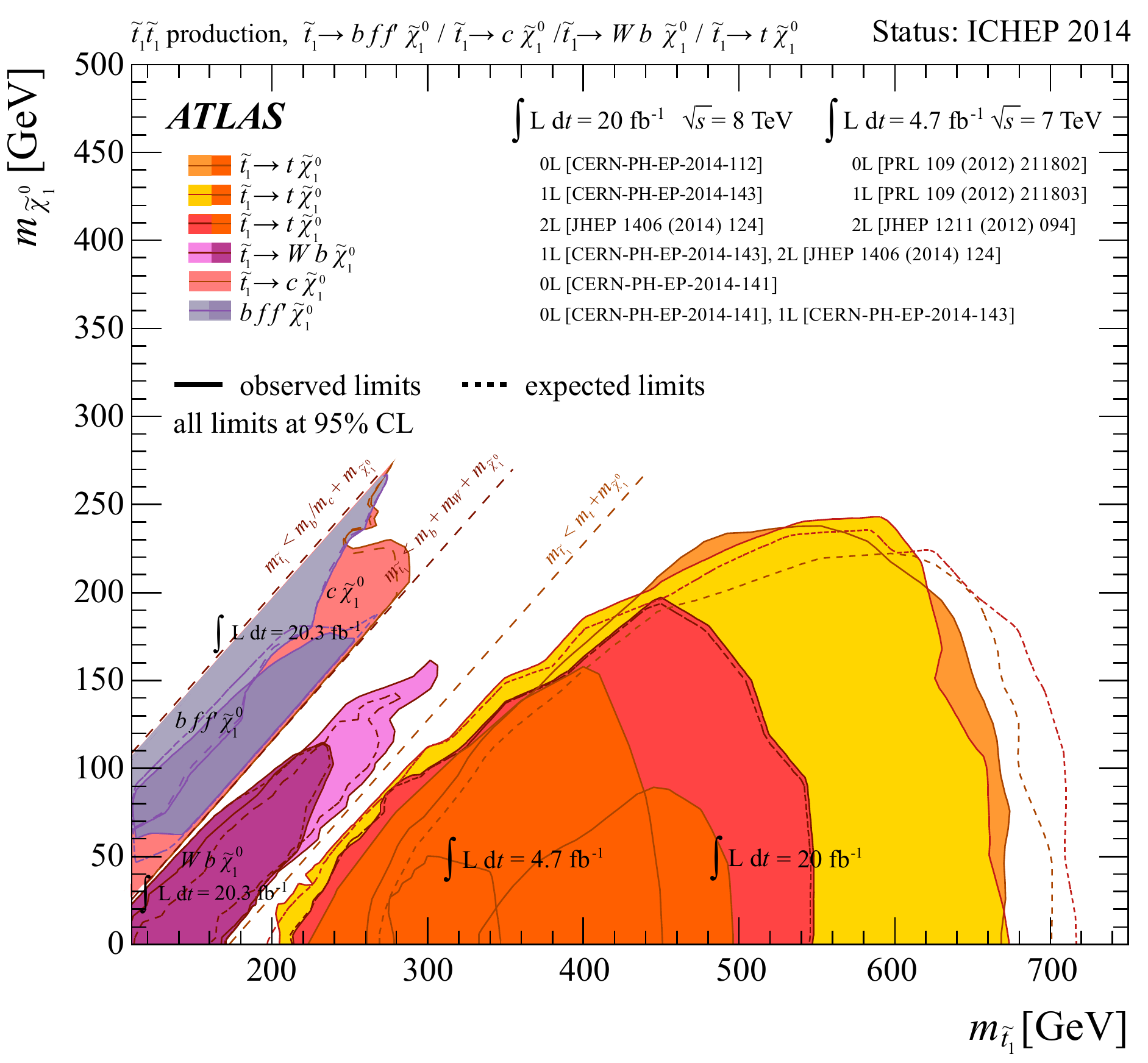}
    \caption{A selection of published limits on the production of third-generation squarks from the ATLAS
      experiment. All limits are given in the form of SMS limits on
      individual production processes (see references in the
      plot). Similar results are obtained from the CMS experiment
      (see e.g.\ Ref.~\cite{Chatrchyan:2013xna}).  
      \textit{(Adapted from Refs.~\cite{Aad:2014bva,Aad:2014kra,Aad:2014qaa,Aad:2014nra,Aad:2012ywa,Aad:2012ywaa,Aad:2012uu}.)}
      \label{fig:SUSY:thirdGenOverview}
}
  \end{center}
\end{figure}

An overview of the currently published search results and limits is
given in \fig{\ref{fig:SUSY:thirdGenOverview}} using examples from the
ATLAS collaboration~\cite{Aad:2014bva,Aad:2014kra,Aad:2014qaa,Aad:2014nra,Aad:2012ywa,Aad:2012ywaa,Aad:2012uu}. 
Similar results are available from
CMS~\cite{Chatrchyan:2013xna}. As expected, the observed sensitivity
is governed by the kinematic regions defined in
\fig{\ref{fig:SUSY:thirdGen:Kinematics}}. The strongest limits reach
up to $m_{\tilde{\tq}_1}>\unit{700}{\GeV}$ for the assumption of the
full strong-production cross section and, more importantly, of 100\%
branching ratio into the given decay. This limit by itself is already
touching the areas which could be considered theoretically interesting
for an elaboration on the natural ability of SUSY to explain the
hierarchy problem of the SM, without unduly fine-tuning the
SUSY-parameters themselves.  
For $m_{\tilde{1}_1}\approx{\mathcal
  O}(\unit{1}{\TeV})$ and higher, the difference between the SUSY
scale and the electroweak scale becomes too large to explain the
hierarchy problem of the SM Higgs mass without additional assumptions.
However, as explained in \sect{\ref{sec:susy:EWthirdsearches:EW}}, the
kinematics of the decays close to the kinematic transitions leads to
vanishing momenta of the visible final-state SM particles, which in
turn yields vanishing exclusions. Therefore, it remains for a global
analysis in a given model to look for candidates of SUSY
implementations which are still in agreement with the allowed low stop
and low neutralino masses. An example of such an analysis is shown in
\sect{\ref{sec:susy:status:Fits}}.

%%%%%%%%%%%%%%%%%%%%%%%%%%%%%%%%%%
%%%%%%%%%%%%%%%%%%%%%%%%%%%%%%%%%
\section{Exotic SUSY Scenarios}
\label{sec:susy:Exoticsearches}

The searches for SUSY described in the previous sections are motivated
by models with a wide range of assumptions. These range from
cosmological constraints (like a neutral stable LSP) over flavour
physics precision measurements (no additional tree-level FCNC) to
aesthetic arguments (fine-tuning). However, none of these are strictly
required to be applicable for a realistic SUSY scenario. For example,
the dark matter in the universe could be explained by something
entirely unrelated to SUSY, while SUSY still could explain electroweak
symmetry breaking and the mass scale of the Higgs boson. It is the aim
of this section to introduce a brief selection of searches in which
the motivation goes beyond the most constraint and most beautiful SUSY
models.\index{dark matter}

The first part of this section deals with charged long-lived heavy
SUSY particles.  The focus of the second part lies on what happens if
the explanation of the dark matter in the universe is omitted from the
motivation of SUSY.  The last part covers scenarios with compressed
mass spectra, i.e.\ scenarios in which the mass difference between the
initially produced SUSY particle and the sum of the decay products is
small. Such scenarios lead to final states with only soft particles
and thus to challenging signatures.

%%%%%%%%%%%%%%%%%%%%%%%%%%%%%%%%%%%%%%%%%%%%%%%%%%%%%%%%%%%%%%%%%%%
\subsection{Searches for Long-Lived SUSY Particles}
\label{sec:susy:Exoticsearches:LongLived}

Long-lived SUSY particles 
%\index{long-lived particles}
can have various origins.  They can occur in models with a very small
mass difference between a charged NLSP and the neutral LSP, in which
case the Lorentz-invariant phase-space volume is small.  Such
scenarios can be realised, for example, in the pMSSM when the
parameter $M_2$ is much smaller than $M_1$ and $\mu$.  Then, the NLSP
and the LSP are wino-like and both have a mass close to $M_2$.
Similarly, the mass difference can
become small when $\mu$ is small compared to $M_1$ and $M_2$, but then
it is typically
a bit larger than in the first case.  Searches for heavy stable
charged particles, which are causing large energy deposits in the
tracking detectors, can be
reinterpreted in SUSY models like the ones mentioned, but this is beyond the scope of this review.

Another potential origin for long-lived SUSY particles are scenarios
with a small coupling of the NLSP to the LSP.  Such a situation can
come about because of a small coupling constant itself, as in the case
of GMSB models where the LSP is the gravitino which couples very
weakly to the NLSP.  Alternatively, the coupling can be small because
of heavy virtual particles in the decay process.  One scenario for
this is ``split SUSY''
\index{SUSY!split SUSY}
~\cite{ArkaniHamed:2004fb, Giudice:2004tc} (see \sect{\ref{sec:susy:theory:breaking}}), where
the gluino and the other gauginos might be accessible at the LHC, but
all scalar SUSY particles are very heavy. Since the decay of the
gluino can only happen via squarks,
it is much suppressed, and the gluino lifetime is long.
The gluino is then forming a
colour-singlet state, a so-called $R$-hadron, which can pass through
the detector and might be stopped by hadronic interactions in one of
the detector components, most likely in the dense calorimeter
material.  Then, after some time, the stopped gluino will decay,
providing a distinct signature of hadronic activity in the calorimeter
without associated tracks pointing to the interaction point. This also
happens in time intervals with no bunch crossings which allows such
events to be triggered with low thresholds on the energy deposit in
the hadronic calorimeter~\cite{Aad:2013gva, Chatrchyan:2012dxa}.
Using this technique, gluino masses up 
to \unit{830}{\GeV} can be excluded for a generic $R$-hadron model with gluino lifetimes
between \power{10}{-6} and \power{10}{4} seconds.

For models with intermediate lifetimes of the NLSP, it
may not reach the outer parts of the detector, but decay inside it.
A possible scenario for such meta-stable particles would be a chargino with the 
decay $\tilde{\chi}^\pm_1\to \Wb^\ast(\to ff')\tilde{\chi}^0_1$.
Since here
the mass difference between LSP and NLSP is small, the momenta of the 
visible decay products are very soft and might not be detectable.
However, if the lifetime is long enough for the chargino
to travel at least through some fraction of the inner tracking detectors, 
the signature of a disappearing track provides sensitivity to such models.
So far, no excess of events with disappearing tracks has been observed at ATLAS or CMS, and a 
lower limit on the mass of meta-stable charginos has been set at \unit{270}{\GeV} 
in one particular SUSY model~\cite{Aad:2013yna}.

%%%%%%%%%%%%%%%%%%%%%%%%%%%%%%%%%%%
%%%%%%%%%%%%%%%%%%%%%%%%%%%%%%%%
\subsection{Searches for $R$-Parity-Violating Models}
\label{sec:susy:Exoticsearches:RPV}

In generic SUSY models with minimal particle content, the SUSY
Lagrangian includes terms that violate conservation of lepton ($L$)
and baryon ($B$) number~\cite{Weinberg:1981wj,Sakai:1981pk}, e.g.\
$$
\frac{1}{2}\rpvlambda L_iL_j\bar{E_k} +
\rpvlambdaP L_iQ_j\bar{D_k} + 
\frac{1}{2}\rpvlambdaPP \bar{U_i}\bar{D_j}\bar{D_k}\, ,
$$
where $L_i$ and $Q_i$ indicate the lepton and quark $\text{SU}(2)$-doublet
superfields, respectively, while $\bar{E_i}$, $\bar{U_i}$ and
$\bar{D_i}$ are the corresponding singlet superfields. The indices
$i$, $j$ and $k$ refer to quark and lepton generations. The Higgs
$\text{SU}(2)$-doublet superfield $H_2$ is the Higgs field that couples to
up-type quarks. The \rpvlambda, \rpvlambdaP{} and \rpvlambdaPP{}
parameters are new Yukawa couplings.

Since the superfields contain both SM particles and SUSY particles,
they e.g.\ couple two SM particles to only one SUSY particle. This means
that, generally, single SUSY particles can be produced, and they can
decay into final states where only SM particles are present.  

In the absence of a protective symmetry, $L$- and $B$-violating terms
may allow proton decay at a rate that is in conflict with the
tight experimental constraints on the proton's
lifetime~\cite{Ahmed:2003sy}.  This difficulty can be avoided by
imposing the conservation of
$R$-parity
\index{R-parity@$R$-parity}
defined as $P_R=(-1)^{3(B-L)+2S}$ as introduced in
\sect{\ref{sec:susy:theory:mssm}}.  However, experimental bounds on
proton decay can also be evaded in $R$-parity-violating
(RPV)
%\index{RPV|see{R-parity violation@$R$-parity violation}}
%\index{R-parity violation|see{R-parity@$R$-parity}}
%\index{R-parity@$R$-parity!violation}
scenarios, as long as the Lagrangian conserves either $L$ or $B$,
because then one of the two steps of the proton decay remains
forbidden: either the production of a single SUSY particle from an up-
and a down-quark in the proton, or the decay of the SUSY particle into
light SM particles, such as leptons and a pion or kaon. Since in RPV
models, single SUSY particles can decay, it means that in such a model
there is no stable LSP, and thus no SUSY candidate for dark matter.

\begin{figure}[t]
  \begin{center}
    \includegraphics[width=0.8\textwidth]{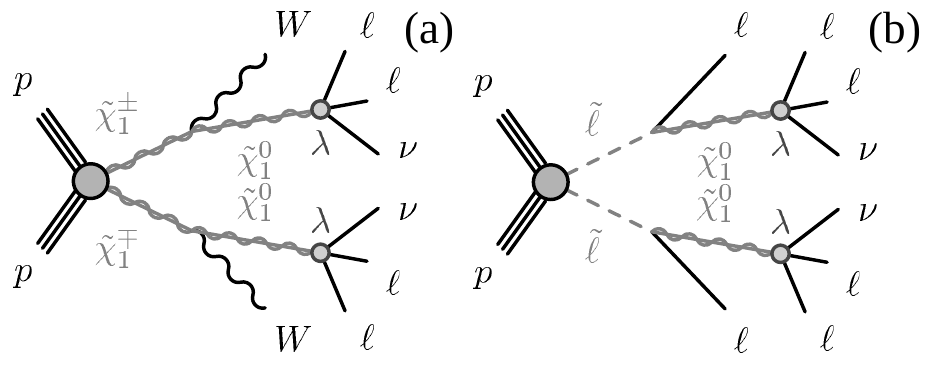}
    \caption{Representative diagrams of SUSY final states in models
      with $R$-parity-violating SUSY and leptonic $R$-parity-violating
      couplings. (a) chargino NLSP; (b) slepton NLSP.
  \label{fig:SUSY:Exotics:RPVdiagrams}
      }
  \end{center}
\end{figure}

Therefore, an obvious way to generate an interesting search scenario
at the LHC that does not contradict proton-lifetime bounds is to just
allow non-zero values of $\rpvlambda$. Then, $B$ remains conserved and
only $L$ can be violated. The SUSY LSP decays into leptons, generating
spectacular signatures consisting of a large number of isolated
charged leptons and \ETmiss only from the neutrinos.  A search along
these lines for final states with at least four isolated leptons,
including electrons, muons and $\tau$ leptons, was performed by the
ATLAS experiment~\cite{Aad:2014iza}. Examples for processes leading to
such final states are shown in
\fig{\ref{fig:SUSY:Exotics:RPVdiagrams}}.

The signal regions selected for such a search are defined by only 3
conditions: First, at least four light leptons or $\tau$ leptons are
required; second, a moderate requirement on $\ETmiss>\unit{50}{\GeV}$
is imposed; and third a veto against the \Zb resonance on same-flavour
opposite-sign leptons is applied.  The veto is applied because the
typical SM backgrounds are rich in \Zb bosons, such as $\Wb\Zb$,
$\Zb\Zb$, $\Zb+$jets and rare processes such as $\Zb\Wb\Wb$,
$\Zb\Zb\Wb$ and $\ttbar\Zb$. No significant excess of the data over
the background is observed, and thus limits are set on the allowed
cross section at the 95\% \CL in SMS scenarios, e.g.\ on topologies
defined by the processes depicted in
\fig{\ref{fig:SUSY:Exotics:RPVdiagrams}}.  Assuming the couplings in
the SMS scenario and 100\% branching ratio in the final state under
study (which, again, is typically not true in any physical model),
mass limits on the NLSP and the LSP can be set. As an example, for a
chargino NLSP up to $m_{\tilde{\chi}^{\pm}_1}>\unit{750}{\GeV}$ can be
excluded for exclusive decays into light leptons, and
$m_{\tilde{\chi}^{\pm}_1}>\unit{550}{\GeV}$ for decays involving
$\tau$ leptons, which allow significantly higher backgrounds from QCD
jets than light leptons. In these searches, the limit does not vanish
for degenerate masses (as in the searches presented in
\sect{\ref{sec:susy:EWthirdsearches}}), because for RPV decays the
search does not depend heavily on the particles emitted between the
NLSP and the LSP. Instead, the sensitivity vanishes for $m_{LSP}\to
\unit{0}{\GeV}$, because then the LSP is highly boosted and its decay
products are collinear and cannot be individually reconstructed
anymore.

%%%%%%%%%%%%%%%%%%%%%%%%%%%%%%%%%%%%%%%%%%%%%%%%%%%%%%%%%%%%%%%%%%%
\subsection{Compressed Spectra}
\label{sec:susy:Exoticsearches:Compressed}

In scenarios where the mass difference of the initially 
produced SUSY particle and the sum of the decay products is small, like e.g.\ in 
the case of compressed SUSY
%\index{compressed SUSY|see{supersymmetry}}
\index{SUSY!compressed SUSY}
~\cite{Martin:2007gf}, the transverse momentum of the final-state objects is in general small.
It is, consequently, very challenging to trigger and select signal events from such models, although---e.g.\ in the case of gluino or squark production--- the production cross section might be very large.
One possibility to get some sensitivity to such models is similar to the search at colliders for direct Dark Matter production: 
%described in \chap{\ref{sec:exo}}, %TODO: reference?
%where the search for directly produced dark matter particles will be discussed:
Initial-state or final-state radiation with high transverse momentum is leading 
to mono-jet or mono-photon signatures, or to signatures with only one $\Wb/\Zb$ boson.
A reinterpretation of these searches in SUSY scenarios provides some sensitivity to compressed mass spectra.

One example is a search for light $\tilde{\tq}_1$ which are almost mass-degenerate 
with the $\tilde{\chi}^0_1$~\cite{Aad:2014nra}.
For such a scenario, the decay via virtual $\Wb^\ast\to ff'$ or the 
CKM-suppressed decay $\tilde{\tq}_1\to \cq\tilde{\chi}^0_1$ are possible (see \sect{\ref{sec:susy:EWthirdsearches:ThirdGen}} for details).
In the latter case the sensitivity can be further improved by distinguishing \cq jets 
from jets with different quark flavour.
The mass ranges in the $m_{\tilde{\tq}_1}$--$m_{\tilde{\chi}^0_1}$ plane that are 
excluded by a search with mono-jet events is shown in \fig{\ref{fig:SUSY:thirdGenOverview}}.

%%%%%%%%%%%%%%%%%%%%%%%%%%%%%%%%%%%%
%%%%%%%%%%%%%%%%%%%%%%%%%%%%%%%
\section{Current Status}
\label{sec:susy:status}
%\index{supersymmetry!status} 
\index{SUSY} 

In the previous sections,
the experimental status of the SUSY searches after LHC Run~1 were
discussed in detail. No statistically significant sign for the
existence of supersymmetric particles has been found. Neither did the
searches for extended Higgs sectors
show any sign of new
physics,
%(\sect{\ref{sec:higgs:additional_higgses}}), %TODO: reference?
nor did the precision measurements in the \Bmeson-meson
system.
% (\chap{\ref{sec:flavour}}). %TODO: reference?
Precision measurements
generally show good agreement with the SM
prediction,
%(\sect{\ref{sec:smew}}), %TODO: reference?
albeit with one potentially
interesting development in the cross-section measurement of \WW
production (\sect{\ref{sec:susy:status:Jamie}}). Thus, limits on SUSY
have been derived from the direct SUSY searches, either in the form of
excluded 2-dimensional hyper-surfaces
in a model-parameter space, or in the form of simplified models. But
what do these limits mean in terms of the understanding of full
physical models of SUSY and \emph{all} their free parameters? How can
we apply these limits, and: Taking all available information from all
experiments together, can we exclude models of SUSY?

It is the aim of this section to explain the status of SUSY. 
In \sect{\ref{sec:susy:status:SMS}} we have already detailed how to use the simplified
model search limits. Section~\ref{sec:susy:status:Fits} then explains
the current status of the global interpretation of SUSY searches and
discusses their remaining viability as an explanation of the hierarchy
problem, dark matter and the unification of
forces.\index{dark matter}\index{unification} Section~\ref{sec:susy:status:Jamie} discusses the
interpretation of the interesting (albeit statistically not
significant) anomaly of the excess in the \WW cross-section
measurement in Run~1 in a SUSY model, and the resulting consequences
for future direct searches for SUSY at the LHC. The section closes
with a summary of the prospects for the LHC Run~2 in
\sect{\ref{sec:susy:status:Prospects}}.

%%%%%%%%%%%%%%%%%%%%%%%%%%%%%%%%%%%%%%%%%%%%%%%%%%%%
\subsection{Global Fits}
\label{sec:susy:status:Fits}

% Not yet published, but interesting and hopefully published very soon:
% https://indico.ific.uv.es/indico/contributionDisplay.py?sessionId=24&contribId=701&confId=2025

If we assume a supersymmetric model with a manageable number of free
parameters, we can estimate which regions of parameter space are
preferred and which are excluded given our current experimental
knowledge. The key observables in such a global parameter fit are the
measured Higgs mass and the relic density attributed to the weakly
interacting dark-matter agent. If we further assume thermal
dark-matter production, the measured relic density requires a sizeable
annihilation cross section of the dark-matter
states~\cite{Jungman:1995df,Bertone:2004pz}. Given the precision of
the Higgs-mass measurement, we also need to include the top mass with
appropriate uncertainties as a model parameter.  Additional
constraints are usually included in our definition of the
supersymmetric model: Custodial symmetry will ensure that we are not
in conflict with electroweak precision data; a flavour symmetry like
minimal flavour violation will keep us away from the rigid flavour
constraints in the $\ensuremath{K}$ and $\ensuremath{B}$ sectors;
leaving out sneutrino dark matter accommodates most of the dark-matter
direct-detection constraints.

In the MSSM,
%\index{supersymmetry!minimal supersymmetric Standard Model} 
\index{SUSY!MSSM} 
the measured Higgs mass gives valuable information on the
radiative corrections mentioned in
\eqn{\eqref{eq:susy:higgsmass}}. The tree-level value as well as the
radiative corrections depend on $\tan \beta$. The leading radiative
correction arises from the top-stop sector and involves the same
logarithm $\log m_{\tilde{\tq}}/\Mt$ that governs the little hierarchy
problem. A relatively heavy Higgs mass requires large stop masses or
at least large stop mixing driven by the trilinear coupling $A_t$
introduced in section~\sect{\ref{sec:susy:theory:breaking}}.  Finally,
the plateau value for $M_h$ mentioned in
\eqn{\eqref{eq:susy:higgsmass}} will only be reached for heavy
additional Higgs states, $\MA > \mathcal{O}(\unit{300}{\GeV})$.

When we rely on gravity mediation or gauge mediation, we tend to
assume that the lightest neutralino is mostly a bino, as discussed in
\sect{\ref{sec:susy:theory:breaking}}. Compared to other compositions
of the lightest neutralino the typical bino annihilation cross
sections are small, which means they predict too large a relic
density. Therefore, scenarios with dominantly bino dark matter have to rely
on a mechanism which enhances their annihilation rate. One way would
be a very light $t$-channel state, like for example the
$\tilde{\tau}$, mediating the annihilation $\tilde{\chi}_1^0
\tilde{\chi}_1^0 \to \tau^+ \tau^-$; alternatively, an LSP mass
matching the mass of a light Higgs $h^0$ or heavy Higgs $A^0$ state
combined with enough of a wino component to couple to the Higgs sector
will allow for a resonant $s$-channel annihilation; we can also
postulate an additional supersymmetric particle roughly within 10\% of
the LSP mass to open co-annihilation processes like, e.g.,
$\tilde{\chi}_1^0 \tilde{f} \to \Zb \ensuremath{f}$ for any Standard
Model fermion
$\ensuremath{f}$~\cite{Griest:1990kh,Mizuta:1992qp,Ellis:1998kh,Ellis:2001nx},
$\tilde{\chi}_1^0 \tilde{\chi}^+ \to
\Wb^+\Zb$~\cite{Binetruy:1983jf,Mizuta:1992qp,Edsjo:1997bg},
$\tilde{\chi}_1^0 \tilde{\chi}_2^0 \to \WW$, or $\tilde{\chi}_1^0
\tilde{\chi}^+ \to \tq\bbarq$.  Finally, in the so-called focus point
region, or, more generally, the so-called well-tempered neutralino
regime of the CMSSM 
%\index{supersymmetry!constrained MSSM} 
\index{SUSY!constrained MSSM} 
the higgsino mass parameter $\mu$ resides close to
the point where it crosses zero and changes sign. This way the light
LSP 
%\index{supersymmetry!lightest supersymmetric particle} 
\index{SUSY!LSP} 
receives a sizeable higgsino fraction, which is one way to avoid
such co-annihilation channels and generate the observed relic density
with the annihilation process $\tilde{\chi}_1^0 \tilde{\chi}_1^0 \to
\WW$.  Similarly, in general models like the pMSSM,
\index{SUSY!pMSSM} 
the LSP can have a
significant wino or higgsino content, enhancing the pair-annihilation
rate sufficiently.

A third class of experimental constraints come from indirect searches
like the branching ratios for \BsToMuMu and $\bq \to \sq \gamma$, and
the anomalous magnetic moment of the muon. The bottom-flavour sector
most effectively constrains parameter regions with small heavy Higgs
masses and large $\tan \beta > \mathcal{O}(30)$. The measured
anomalous magnetic moment of the muon requires light sleptons and
light electroweak gauginos.  In combination with other measurements
this last requirement can be hard to fulfill, which means that the
measured anomalous magnetic moment often adds an offset to the quality
of the global fits.

Direct LHC searches enter the global fits as limits on the masses (and
couplings) of the new states. The LHC limits on the weakly interacting
sector are relatively mild. In contrast, the weakly interacting
gaugino and slepton sectors are highly sensitive to dark matter
observables and, if light, to limits from direct LEP searches. The
latter essentially require all new states coupling to the weak gauge
sector to have masses above half of the LEP centre-of-mass energy. In
the strongly interacting sector the observation of a Higgs boson at a
mass far beyond its tree-level bound $M_h < \MZ$ but in agreement with
the loop-corrected value of \eqn{\eqref{eq:susy:higgsmass}} points to
relatively large squark and gluino masses.  This is in general
agreement with the non-observation of SUSY in direct searches, so the
direct LHC searches will not add much information to a global
analysis.

Other constraints, e.g.\ from possible signals in the search for
emissions of neutralino annihilation in the universe, do not enter the
global fits. The reason is that possible signals extracted either from
$\gamma$-ray data~\cite{Ackermann:2012qk,Weniger:2012tx} or from
positron data~\cite{Adriani:2013uda}
collected with satellite experiments are subject to significant
systematic uncertainties on the astrophysical background, and thus is
not necessarily a sign of physics beyond the SM.

\begin{figure}[t]
  \begin{center}
    \includegraphics[width=0.95\textwidth]{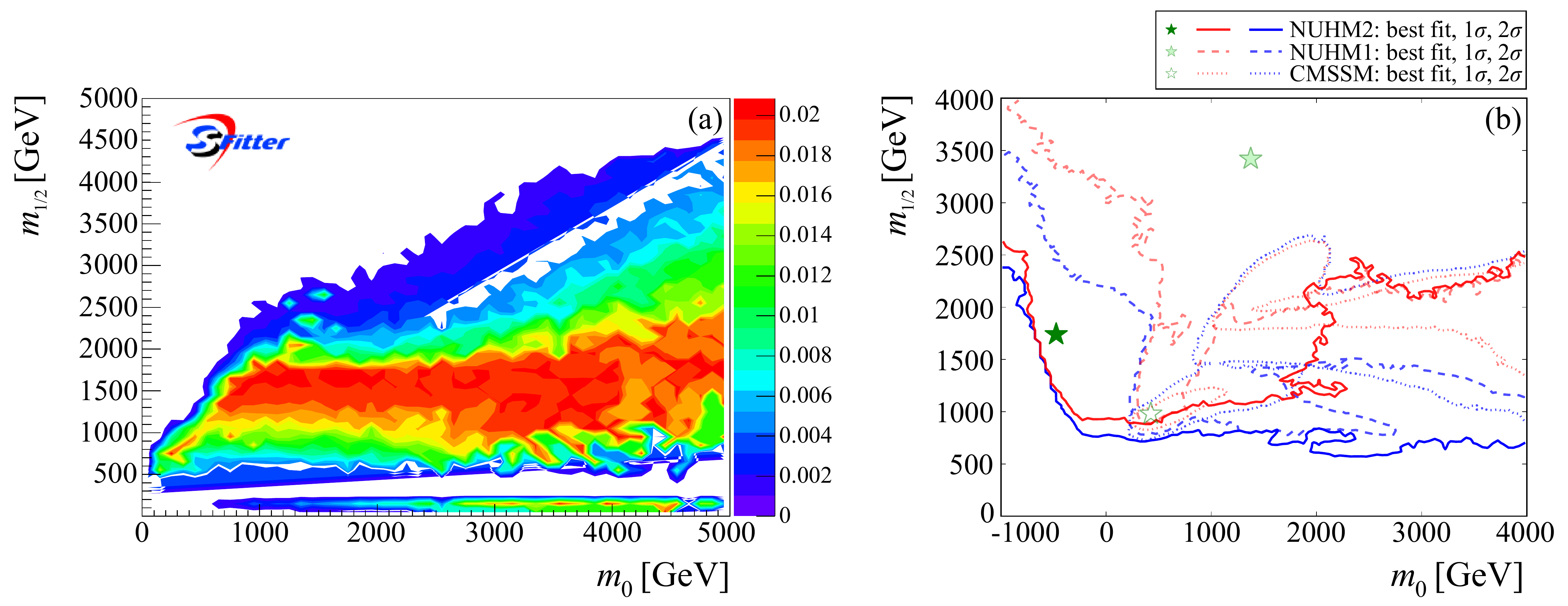}
  \caption{Fit results in the CMSSM.  (a) Frequentist profile
    likelihood in the $m_0$--$m_{1/2}$
    plane.  
    (b) Comparison of the
    CMSSM and the NUHM.
    \textit{(Adapted from
      Refs.~\cite{Henrot-Versille:2013yma,Buchmueller:2014yva}.)}
  \label{fig:SUSY:fits:CMSSM}
}
%A comparison of the predicted direct detection
%    cross-section of dark matter~\cite{Roszkowski:2014wqa} with the
%    predictions for future experiments like
%    XENON1T~\cite{Aprile:2012zx} is shown on the lower left. 
  \end{center}
\end{figure}

Without discussing the technical details, e.g.\ on the differences in
the Bayesian and frequentist approaches to supersymmetric parameter
analyses, we will start from the relatively simple CMSSM case and then
move towards more general models described in
\sect{\ref{sec:susy:theory:breaking}}.  As described above, the main
constraints on the supersymmetric parameter space will come from a
combination of the Higgs-mass prediction with and the properties of
the weakly interacting dark-matter agent. The more constrained the
model is, the more clearly we can then identify preferred parameter
regions. In particular the dark matter constraint strongly reduces the
number of allowed parameter regions, for example in the
CMSSM~\cite{Strege:2012bt,Cohen:2013kna,Henrot-Versille:2013yma,Roszkowski:2014wqa,Bechtle:2014yna,Buchmueller:2014yva}:
In \fig{\ref{fig:SUSY:fits:CMSSM}}(a) we show the profile likelihood
in the two mass parameters $m_0$ and $m_{1/2}$, where the latter
largely determines the mass of the LSP. The LHC limits are not
included in this figure, so a narrow strip at low but constant
$m_{1/2}$ values indicates $h^0$-pole annihilation. Such low universal
gaugino masses are ruled out by gluino searches at the LHC. The
strongly correlated region for $m_{1/2} < \unit{1}{\TeV}$ corresponds
to stau co-annihilation with a light bino. The absence of the 
corresponding narrow line-shaped region we see that the focus point region
with a light higgsino LSP is essentially ruled out.  The bulk region
around $m_{1/2} = \unit{1.5}{\TeV}$ relies on the heavy Higgs-pole
annihilation~\cite{Henrot-Versille:2013yma}.  It is merged with the
so-called well-tempered neutralino scenario, both dominating the
allowed CMSSM parameter space.

In this CMSSM parameter study, large values $\tan \beta > 40$ are
preferred in order to accommodate the relatively large Higgs-mass
value.  As expected, the allowed parameter regions by the Higgs mass
and the relic density touch the LHC exclusion for low $m_{1/2}$, but
the preferred regions are beyond the sensitivity of direct LHC
searches (see \fig{\ref{fig:SUSY:ATLAS_CMSSM}}). In
\fig{\ref{fig:SUSY:fits:CMSSM}}(b) a similar result is shown, now
based on the NUHM 
%\index{supersymmetry!non-universal Higgs mass ansatz} 
setup introduced
in~\sect{\ref{sec:susy:theory:breaking}}.  While the best-fit value
moves around in the $m_0$ vs $m_{1/2}$ plane, 
owing to a relatively flat profile
likelihood profile, the qualitative result hardly
changes~\cite{Buchmueller:2014yva}.

\begin{figure}[t]
  \begin{center}
    \includegraphics[width=0.95\textwidth]{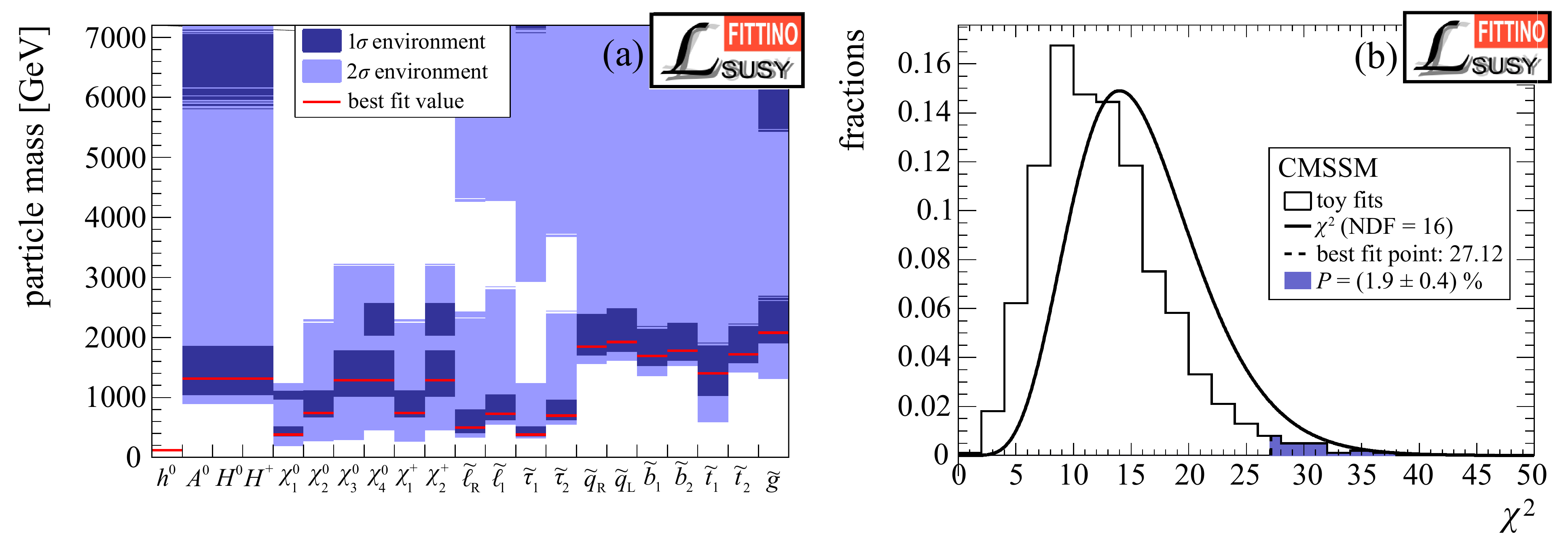}
  \caption{(a) Mass spectrum for the best-fit point in the CMSSM
    including error bars. 
    (b) Calculation of the CMSSM $p$-value 
    below 5\% \CL.
    \textit{(Adapted from Ref.~\cite{Bechtle:2014yna}.)}
%    The results of
%    the fits from Fig.~\ref{fig:SUSY:fits:CMSSM} expressed as mass
%    spectra of the physical SUSY particles and Higgs bosons. On the
%    left, the allowed distribution and best fit values for the CMSSM
%    are shown~\cite{Bechtle:2012zk,Bechtle:2013mda}. 
  \label{fig:SUSY:fits:CMSSMmasses}}
  \end{center}
\end{figure}

An interesting question is how well the CMSSM---as one of the most
constrained models for the supersymmetric mass spectrum---agrees with
all available data in a comprehensive statistical analysis.  In
\fig{\ref{fig:SUSY:fits:CMSSMmasses}}(a) the mass spectrum of the
global best-fit parameter point of the MSSM is
shown~\cite{Bechtle:2014yna}.  As expected, gluinos and light-flavour
squarks are pushed to high masses. The allowed heavy Higgs states
cover a wide range, but also prefer to be heavy. Most of the weakly
interacting new states prefer to be light, but are not necessarily
accessible in direct production in Run~1. The dark-matter annihilation
in this best-fit point proceeds through stau co-annihilation.  In
\fig{\ref{fig:SUSY:fits:CMSSMmasses}}~(b) the statistical analysis of
the CMSSM and the corresponding log-likelihoods are shown. The smooth
curve shows the naive expectation of the distribution of
log-likelihoods, while the histogram shows the actual result from toy
experiments. The problem with all CMSSM fits is that slepton and
squark masses are linked through the universal scalar mass assumption.
Experimentally, the large measured value of the anomalous magnetic
moment of the muon prefers light sleptons, while the stronger
Higgs-mass constraint prefers heavy squarks.  Correspondingly, the
overall $\chis$ equivalent shows that the quality of the fit is poor
and the corresponding $p$-value is low. In essence, the CMSSM is ruled
out at more than 95\%~\CL~\cite{Bechtle:2014yna}.

\begin{figure}[t]
  \begin{center}
    \includegraphics[width=0.95\textwidth]{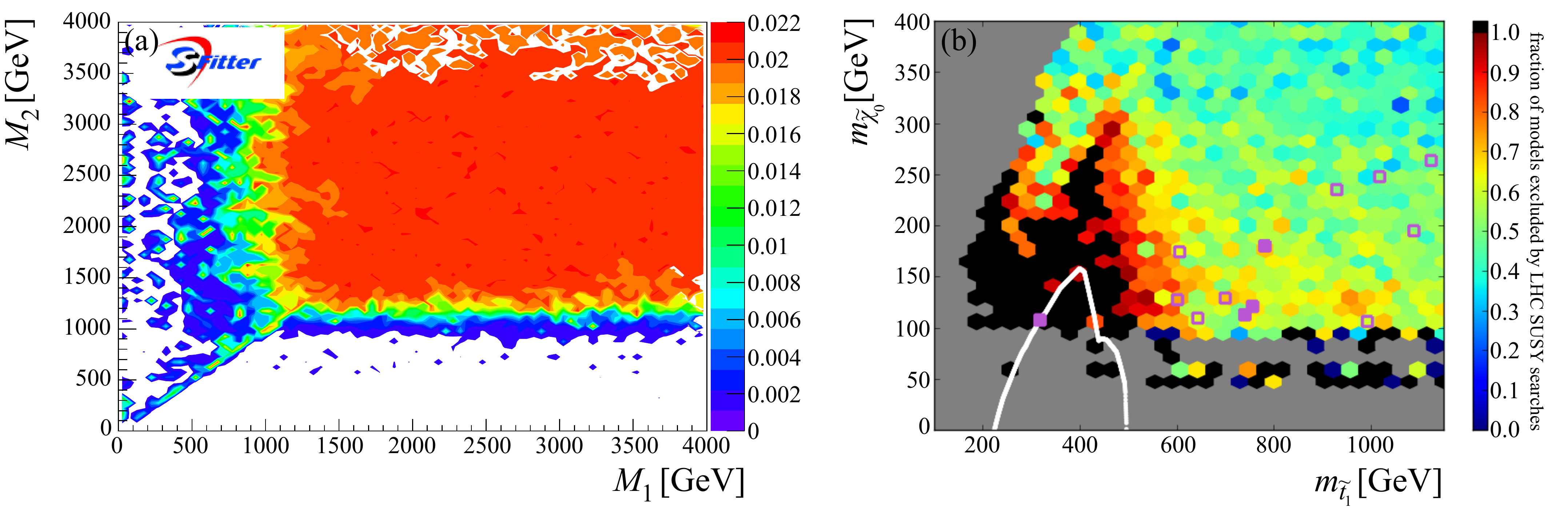}
  \caption{Fit results in the pMSSM. (a) Frequentist profile
    likelihood in the bino-wino mass plane before applying LHC
    results. (b) Fraction of excluded pMSSM models in the
    LSP-$\tilde{\tq}_1$ mass plane.  Overlaid is a limit from a
    simplified-model interpretation.  \textit{(Adapted from
      Refs.~\cite{Henrot-Versille:2013yma,CahillRowley:2012kx}.)}
    \label{fig:SUSY:pMSSM_LSP_Gluino}
  }
  \end{center}
\end{figure}

This tension in the CMSSM suggests to look at significantly more
general, but still predictive parametrisations of SUSY. The biggest
challenge in pMSSM 
\index{SUSY!pMSSM} 
parameter studies is to effectively scan the
19-dimensional parameter space~\cite{CahillRowley:2012kx,
    Boehm:2013qva,Henrot-Versille:2013yma}.  In
\fig{\ref{fig:SUSY:pMSSM_LSP_Gluino}} we show the 2-dimensional
profile likelihood in the bino--wino mass
plane~\cite{Henrot-Versille:2013yma}. As for the CMSSM, the dark
matter constraint determines the main features of the allowed
parameter range in the weakly interacting sector. The strongly
correlated diagonal strip for $M_1 \sim M_2 \sim 1$~TeV corresponds to
stau co-annihilation; the individual points around $M_1 \sim 63$~GeV
will at sufficiently large statistics form a line resulting from
$h^0$-pole annihilation; the bulk region for large $M_1$ and $M_2$
receives contributions from heavy Higgs-pole annihilation, or, more
generally the well-tempered neutralino scenario including chargino
co-annihilation. There exists allowed parameter space for stop
co-annihilation, not covered by this specific parameter analysis.

Based the Higgs-mass constraints a pMSSM analysis can for example map
out the preferred parameter space of the extended Higgs
sector~\cite{Dumont:2013npa}.  One of the most interesting parameter
correlation in combination with dark matter then becomes the bino
mass vs the lightest stop mass. The former represents the dark matter
constraints while the latter probes the Higgs-mass constraints.  In
\fig{\ref{fig:SUSY:pMSSM_LSP_Gluino}} the fraction of pMSSM parameter
points excluded by the LHC searches is shown. Obviously, parameter
points with low masses are more often excluded by the searches
discussed in \sect{\ref{sec:susy:EWthirdsearches}}. Comparing the
contour of constant fraction of rejected points to the contour from a
simplified-model interpretation, we observe differences.  The reason
lies in the assumed decays. On the one hand, actual branching ratios
smaller than 100\% result in a weaker limit or fewer excluded points
than for the simplified model. 
\index{SUSY!simplified models} 
On the other hand, decay modes not
governed by the simplified model can have an enhanced sensitivity and
consequently lead to stronger limits or more excluded points. This
difference illustrates the problem with simplified model
interpretations. Their limits can be applied to all specific SUSY
models, but the actual exclusion lines are not the same in complete
models. There, the limit might be stronger or weaker, depending on its
mass and coupling structure. Note that results based on the fraction
of rejected points must be interpreted carefully. First, a scan of a
19-dimensional parameter space is computationally very demanding, so
we are not guaranteed that we do not miss allowed points for example
at lower masses. An example is the disrupted line of $h^0$-pole
annihilation in \fig{\ref{fig:SUSY:pMSSM_LSP_Gluino}}.  Second,
the statistical interpretation of the fraction of excluded points
includes an ad-hoc Bayesian measure.

The main conclusion of the supersymmetric parameter studies is that
including the Higgs measurements of Run~1, the available data on dark
matter and indirect constraints, as well as the LHC exclusion bounds
the MSSM is not under significant pressure -- in strong contrast to
the more popular constraint models as the CMSSM.  It still offers an
explanation for dark matter, for the hierarchy problem, and for
approximate gauge coupling unification.

%%%%%%%%%%%%%%%%%%%%%%%%%%%%%%%%%%%%%%%%%%%%%%%%%%%%
\subsection{Experimental Anomalies}
\label{sec:susy:status:Jamie}

From any experiment based on statistical analyses
and including significant systematic complexity we should always 
expect experimental anomalies which lend themselves to new physics interpretations.
The crucial test of any anomaly pointing, for example, to
physics beyond the Standard Model will be if it becomes more
significant when more data are included. A few such anomalies exist at
the end of Run~1 at the LHC. They hardly contribute to the global
fits, but they maybe interesting to follow up when LHC resumes data
taking in 2015.

\begin{figure}[t]
  \begin{center}
    \includegraphics[width=0.65\textwidth]{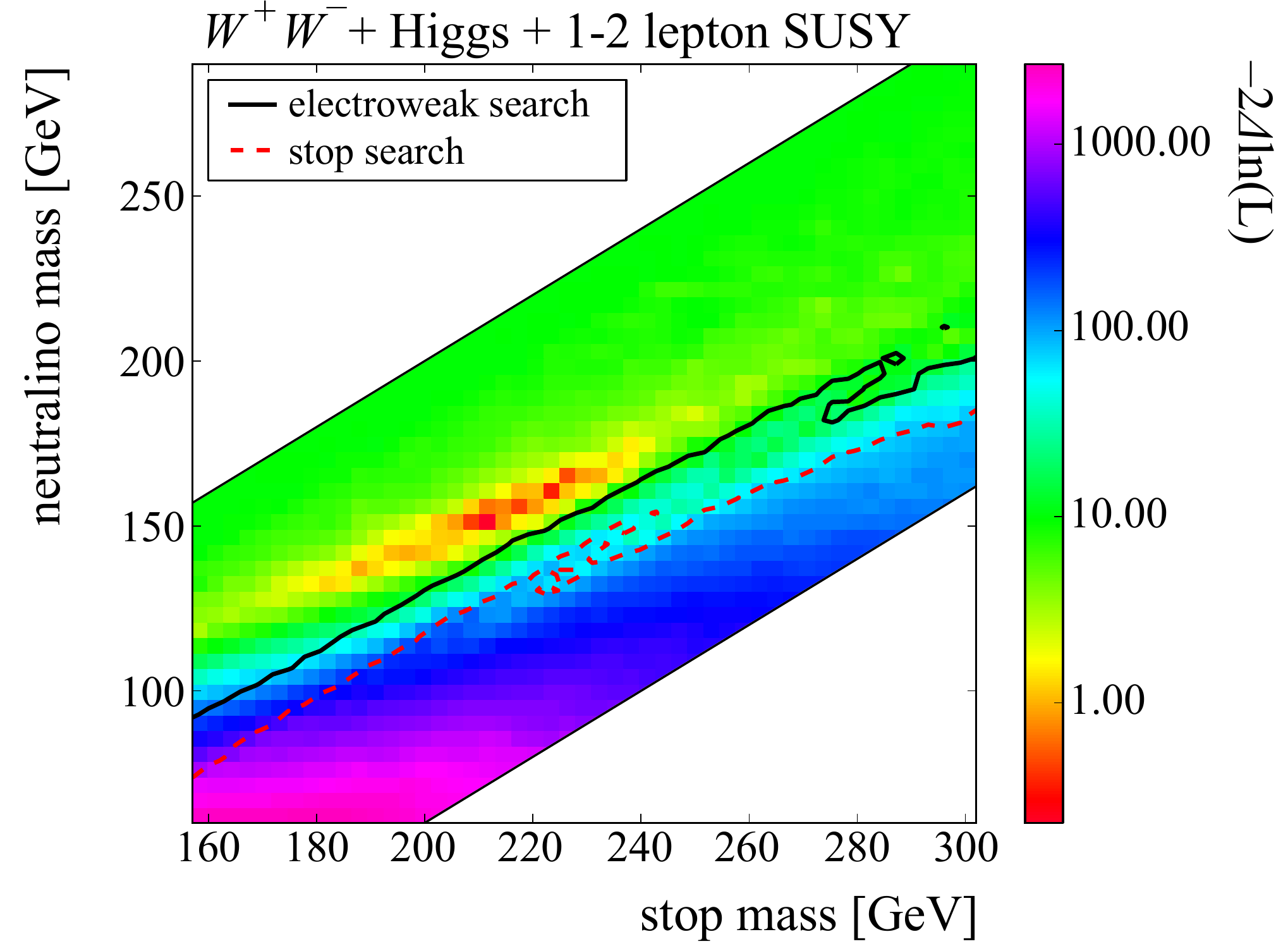} 
  \caption{Two-dimensional likelihood for the relevant mass
    parameters in the decay chain shown in
    \eqn{\eqref{eq:SUSY:anomaly}}. The analysis was performed on ATLAS
    and CMS \WW measurements.  
    \textit{(Adapted from Ref.~\cite{Kim:2014eva}.)}
  \label{fig:SUSY:anomaly}
}
  \end{center}
\end{figure}

Already the \unit{7}{\TeV} analyses of \WW production in
ATLAS~\cite{ATLAS:2012mec} and CMS~\cite{Chatrchyan:2013yaa} showed
only modest agreement between the data and Standard Model
simulations. General problems with this channel, both in the total
rate and in some of the distributions, have been known for a long
time~\cite{Curtin:2012nn}.  The phenomenological analyses of
Refs.~\cite{Kim:2014eva,Curtin:2014zua} also includes some CMS
\unit{8}{\TeV} results in the same
channel~\cite{Chatrchyan:2013oev}. The experimental signature that can
be interpreted as Standard Model background plus supersymmetry signal,
is dileptons plus missing transverse momentum.
The signature includes a jet veto to reject the top-quark
pair production background.  The assumed signal process is
%
%\begin{linenomath}
\begin{equation}
\pp 
\to \tilde{\tq}_1 \tilde{\tq}_1^*
\to (\tilde{\chi}_1^+ \bq) \; (\tilde{\chi}_1^- \bbarq) 
\to (\tilde{\chi}_1^0 \ell^+ \nu \bq) \; (\tilde{\chi}_1^0 \ell^- \bar{\nu} \bbarq) \, .
\label{eq:SUSY:anomaly}
\end{equation}
%\end{linenomath}
%
This production and decay process is similar to the cascade decay with its 
dilepton edge discussed in \sect{\ref{sec:susy:theory:signatures}}.
However, rather than through a slepton, the chargino decay proceeds through an
off-shell \Wb boson.  The final state differs from the \WW
background only in the 
the additional existence of two bottom-quark jets.
However, as long as the masses of all new particles involved are small
and the mass difference does not exceed
$m_{\tilde{\tq}} - m_{\tilde{\chi}^\pm} \sim \unit{7}{\GeV}\, ,$
these jets will be too soft to be
observable in the general QCD activity of the events. If the
mass difference between the stop and the chargino is fixed, the two free
parameters in this analysis are the stop mass and the LSP mass. 
Figure~\ref{fig:SUSY:anomaly} shows the likelihood of a scan of the
two-dimensional mass plane. It is dominated by the event count in the
signal regions of the three LHC analyses mentioned above. The stop
branching ratio to the final state of \eqn{\eqref{eq:SUSY:anomaly}} is
assumed to be 100\%.  The strong correlation follows constant mass
differences between the stop, the chargino, and the LSP. The best-fit
point is
%
%\begin{linenomath}
\begin{equation*}
    m_{\tilde{\tq}} = 212^{+35}_{-40}~\GeV
  \qquad \text{and} \qquad 
    m_{\tilde{\chi}^0_1} = 150^{+30}_{-20}~\GeV \, .
\end{equation*}
%\end{linenomath}
%
The estimated significance of this single anomaly is around 2.6
standard deviations and can be increased by including tri-lepton
signals~\cite{Kim:2014eva}. The final individual and combined
significances will have to be determined in a full experimental
analysis. 

In a situation like this it is crucial to ensure that the potential
signal for new physics is not ruled out by any other LHC searches. For
example, searches for direct electroweak gaugino production in the
2-lepton plus missing energy channel contribute to the more
constraining of the two exclusion limits shown in
\fig{\ref{fig:SUSY:anomaly}}. The second exclusion limit arises from
direct stop searches including bottom-quark jets.

As discussed in \sect{\ref{sec:susy:theory:signatures}} the invariant
mass distribution of two leptons, $m_{\ell \ell}$, is a standard
signature for (supersymmetric) cascade decays. Interestingly, there exist indications
for such an anomaly in two opposite-sign same-flavor leptons combined
with missing transverse momentum both in ATLAS~\cite{Aad:2015wqa} and in CMS~\cite{Khachatryan:2015lwa, Allanach:2014gsa}. 
While the ATLAS excess of about $3\,\sigma$ appears directly on the \Zo resonance at $m_{\ell\ell}=\unit{91.2}{\GeV}$, the CMS excess with a significance of $2.6\,\sigma$ appears as an edge-like structure at $m_{\ell\ell}$ below the \Zo resonance. 
As discussed in \sect{\ref{sec:susy:theory:signatures}} the main
backgrounds can be controlled by opposite-sign opposite-flavor
subtraction. The edge of the corresponding distribution appears around
\unit{80}{\GeV}. From this information we can compute the hypersphere
of supersymmetric mass values following \eqn{\eqref{eq:SUSY:edge}}.

A sizeable number of events under such an edge points to a strongly
interacting hard scattering process.  To relate a possible LHC signal
to the direct production of light-flavor squarks and gluinos, we have
to take into account the limits from generic searches for jets plus
missing energy, which only allow heavy light-flavour squarks and
gluinos.  In comparison, the production rates of third-generation
squark pairs are expected to be considerably smaller: First, the
light-flavour rates include an additional factor 8 for experimentally
indistinguishable squarks, as listed in
\tab{\ref{tab:susy:particles}}; second, for light flavours the
$t$-channel gluino exchange contributes up to half of the cross
section, while it is strongly suppressed for sbottoms and non-existent
for stops. To avoid experimental constraints we could link an observed
$m_{\ell \ell}$ edge to the production of relatively light sbottom
pairs, $\pp \to \tilde{\bq} \tilde{\bq}^*$. The typical supersymmetric
parameter range corresponding to such an explanation is similar to the
masses required for the $WW$ anomaly discussed above.

Of course, these anomalies could have a number of explanations that are not
related to supersymmetry. However, it shows that at the end of Run~1,
there are open questions in LHC data which have to be watched in the
upcoming runs.

%%%%%%%%%%%%%%%%%%%%%%%%%%%%%%%%%%%%%%%%%%%%%%%%%%%%
\subsection{Prospects for LHC Run~2}
\label{sec:susy:status:Prospects}

As was shown in \sect{\ref{sec:susy:status:Fits}}, the CMSSM 
\index{SUSY!constrained MSSM} 
as the
prime example for a constraint SUSY model is at the verge of being
completely excluded.  However, the exclusion of a single highly
constraint model in a global analysis must not be taken as a hint
against the realisation of a more general SUSY model. Therefore, also
in LHC Run~2, SUSY searches will be of high importance. As motivated
in the pMSSM 
\index{SUSY!pMSSM} 
studies in the previous section, special attention will
be given to searches for electroweak SUSY partners, sleptons and
third-generation squarks. If these remain light (and thus more easily
accessible at the LHC), a naturalness problem in the SUSY sector
stemming from the difference between the electroweak scale and the
mass scale of the SUSY particles can be avoided. Thus gauginos,
sleptons and third-generation squarks are prime objectives for the
LHC.
 
\begin{figure}[t]
  \begin{center}
    \includegraphics[width=0.95\textwidth]{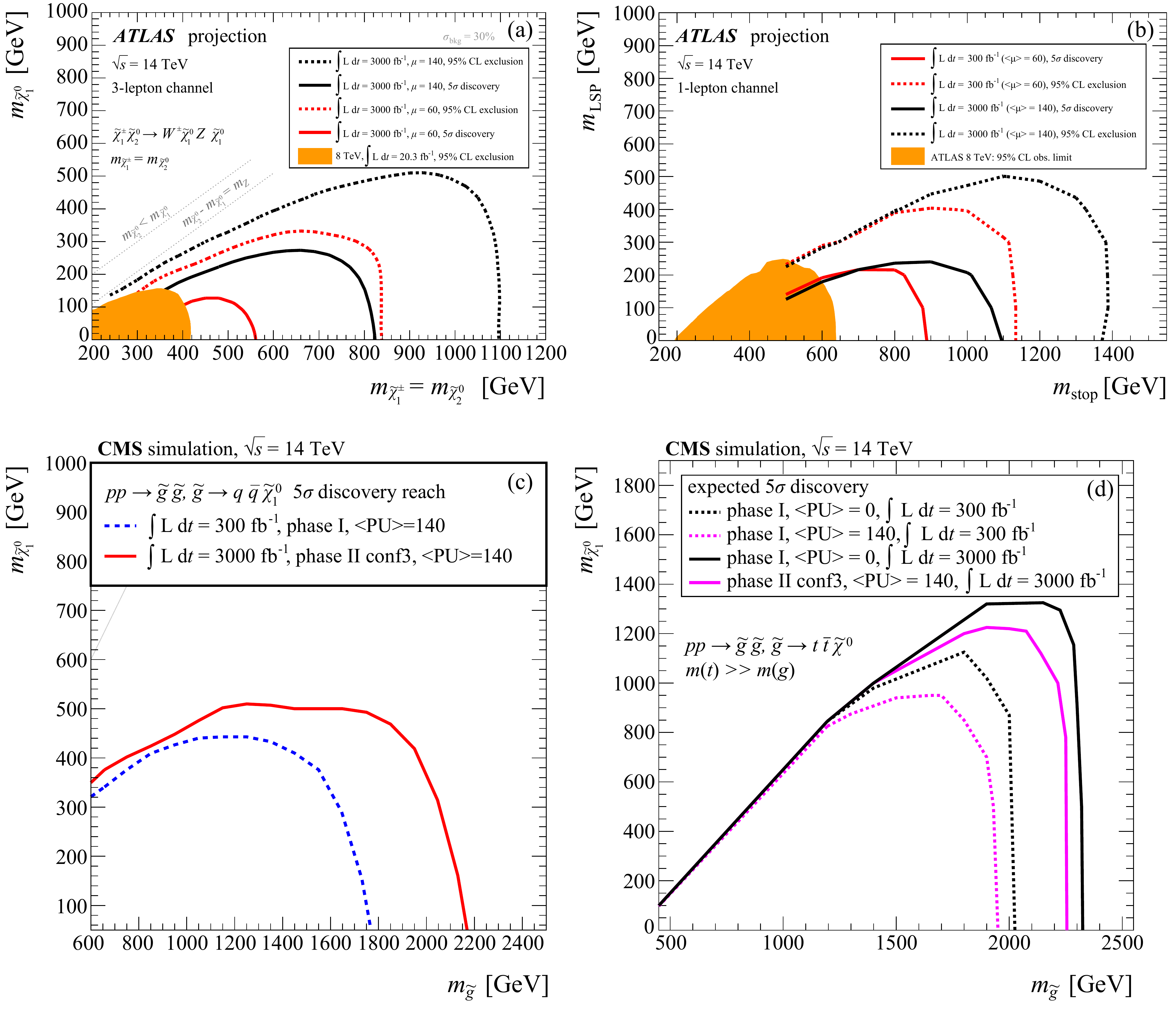}
    \caption{ 
      (a) The expected 95\% \CL exclusion (dashed lines) and discovery
      contours (solid lines) for LHC luminosity scenarios with
      \unit{300}{\invfb} (red) and \unit{3000}{\invfb} (black) in the
      $\tilde{\chi}^{\pm}_{1}/\tilde{\chi}^0_2$--$\tilde{\chi}^0_1$
      mass plane for a simplified model with 100\% branching ratio of
      $\tilde{\chi}^{\pm}_1/\tilde{\chi}^0_2\to
      (W/Z)\tilde{\chi}^0_1$~\cite{ATL-PHYS-PUB-2014-010}. The
      measured \unit{8}{\TeV} exclusion contour is displayed as the
      orange area~\cite{Aad:2014vma}. 
      (b) Same as (a) but the $\tilde{\tq}_1$--$\tilde{\chi}^0_1$ mass plane assuming the decay $\tilde{\tq}_1\to \tq\tilde{\chi}^0_1$ with a branching ratio of 100\% in the 1-lepton decay channel~\cite{ATL-PHYS-PUB-2013-011}. In addition, the observed excluded region from Ref.~\cite{Aad:2014kra} is shown.
      (c) The $5\,\sigma$ discovery reach for LHC luminosity scenarios with \unit{300}{\invfb} (blue dashed line)
      and \unit{3000}{\invfb} (red solid line) in the $\tilde{g}$--$\tilde{\chi}^0_1$ plane for an assumed 100 \% branching ratio of $\tilde{g}\to q\bar{q}\tilde{\chi}^0_1$~\cite{CMS:2013xfa}.
      (d) Same as (c) but for a 100 \% branching ratio of $\tilde{g}\to q\bar{t}\tilde{\chi}^0_1$, shown for no (black) and realistic (purple) pile-up conditions~\cite{CMS:2013xfa}.
      \textit{(Adapted from Refs.~\cite{ATL-PHYS-PUB-2013-011,Aad:2014kra,CMS:2013xfa}.)}
        \label{fig:SUSY:prospects}
      }
  \end{center}
\end{figure}

As an example of studies for gaugino and top squark searches at higher
luminosity, results from
ATLAS~\cite{ATL-PHYS-PUB-2014-010,ATL-PHYS-PUB-2013-011} are
shown in \fig{\ref{fig:SUSY:prospects}}. The existing observed limits
derived from the non-observation of
SUSY~\cite{Aad:2014vma,Aad:2014kra} are compared to the expected reach
at the LHC with an ultimate integrated luminosity of
\unit{300}{\invfb} and at the
HL-LHC
%~\ref{sec:MachineSectionExplainingTheHL-LHC} %TODO: reference?
with an ultimate
integrated luminosity of \unit{3}{\invab}. Similar projections exist
for CMS~\cite{CMS:2013xfa}. The plots show both the reach for
exclusion (dashed) at 95\% \CL and for a $5\,\sigma$ discovery. It is
obvious that already the discovery reach at ${\mathcal
  L}=\unit{300}{\invfb}$ extends significantly beyond the current
exclusion, and thus great discoveries might be expected.
However, if the mass scales of SUSY particles in all viable
models could be pushed beyond the ${\mathcal O}(\TeV)$ region, SUSY
would completely loose its attractiveness as an explanation of the
hierarchy problem.

The LHC prospects for searches after gluino pair production are also shown in \fig{\ref{fig:SUSY:prospects}}, both for scenarios with 100\% branching ratio of $\tilde{g}\to q\bar{q}\tilde{\chi}^0_1$ and $\tilde{g}\to t\bar{t}\tilde{\chi}^0_1$~\cite{CMS:2013xfa}.
In both scenarios the expected discovery reach is extended to 2 \TeV and beyond.

On the other hand, the exclusion and discovery potential is still
expected to be small for specific mass cases, e.g.\ for
$m_{\tilde{\chi}^0_2}-m_{\tilde{\chi}^0_1}\approx \MZ$ or
$m_{\tilde{\tq}_1}-\Mt\approx m_{\tilde{\chi}^0_1}$. In such a case, the
momentum of the SM decay particles approaches 0. Therefore, the
trigger and selection efficiencies become small and the discovery or
exclusion potential can be weakened. The same mechanism applies also to
other searches. Thus, it can be expected that also in the unfortunate
case of a non-discovery of SUSY in LHC Run~2 and beyond, the fate of
SUSY will not be decided by the achievable exclusions directly, but,
if at all, by a global analysis looking for SUSY models which fulfill
at the same time all constraints from direct searches and all indirect
constraints from other sensitive measurements, as outlined in
\sect{\ref{sec:susy:status:Fits}}.

%%%%%%%%%%%%%%%%%%%%%%%%%%%%%%%
%%%%%%%%%%%%%%%%%%%%%%%%%%%%%%%%%%%%
\section{Summary}\label{sec:susy:summary}

Although SUSY still is the best-motivated extension of the 
Standard Model, the LHC Run~1 did not provide us with a clear verdict 
on its existence. The MSSM still provides explanations
for the hierarchy problem and the unification of the forces, as well
as for the observation of the SM-like Higgs boson, the cold dark matter in
the universe, and for experimental anomalies such as the measurement of the
anomalous magnetic moment of the muon $(g-2)_{\mu}$.  All that can be
achieved at a mass scale of the lightest SUSY particles of at most a
few hundred \GeV.  The observed Higgs mass, for example, or dark-matter
searches could have ruled out the MSSM, but they did not.

No statistically significant hint for new physics
has been found in Run~1. The combination of the Higgs-mass
measurement and the direct mass limits on SUSY particles on the one
hand, and observables like $(g-2)_{\mu}$ on the other hand have
excluded strongly constrained models like the CMSSM at the
95\% \CL level. 
This is a great success of combined
experimental studies at the LHC, at \Bmeson factories, at the \Tevatron, LEP,
from low-energy precision physics, and from cosmology. With the increased
energy and luminosity of LHC Run~2, we can expect definite answers on new
physics models which include strongly interacting states at the
few-\TeV\ scale, including the MSSM.

\bibliography{lhc_susy}{}

\end{document}